\documentclass[a4paper]{report}
\usepackage[utf8]{inputenc}
\usepackage[T1]{fontenc}
\usepackage{RJournal}
\usepackage{amsmath,amssymb,array}
\usepackage{booktabs}

\usepackage{multirow}

\begin{document}

\sectionhead{Contributed research article}
\volume{XX}
\volnumber{YY}
\year{20ZZ}
\month{AAAA}

\begin{article}
\title{The Landscape of R Packages for Automated Exploratory Data Analysis}
\author{by Mateusz Staniak and Przemysław Biecek}

\maketitle

\abstract{
 The increasing availability of large but noisy data sets with a large number of heterogeneous variables leads to the increasing interest in the automation of common tasks for data analysis. The most time-consuming part of this process is the Exploratory Data Analysis, crucial for better domain understanding, data cleaning, data validation, and feature engineering. 

 There is a growing number of libraries that attempt to automate some of the typical Exploratory Data Analysis tasks to make the search for new insights easier and faster. In this paper, we present a systematic review of existing tools for Automated Exploratory Data Analysis (autoEDA). We explore the features of fifteen popular R packages to identify the parts of the analysis that can be effectively automated with the current tools and to point out new directions for further autoEDA development.
}

\section{Introduction}

With the advent of tools for automated model training (autoML), building predictive models is becoming easier, more accessible and faster than ever. 
Tools for R such as mlrMBO \citep{mlrMBO}, parsnip \citep{parsnip}; tools for python such as TPOT \citep{tpot}, auto-sklearn \citep{autosklearn}, autoKeras \citep{autokeras} or tools for other languages such as H2O Driverless AI \citep{h2o, h2obook} and autoWeka \citep{autoweka} supports fully- or semi-automated feature engineering and selection, model tuning and training of predictive models. 

\begin{figure}
    \centering
    \includegraphics[width=\textwidth]{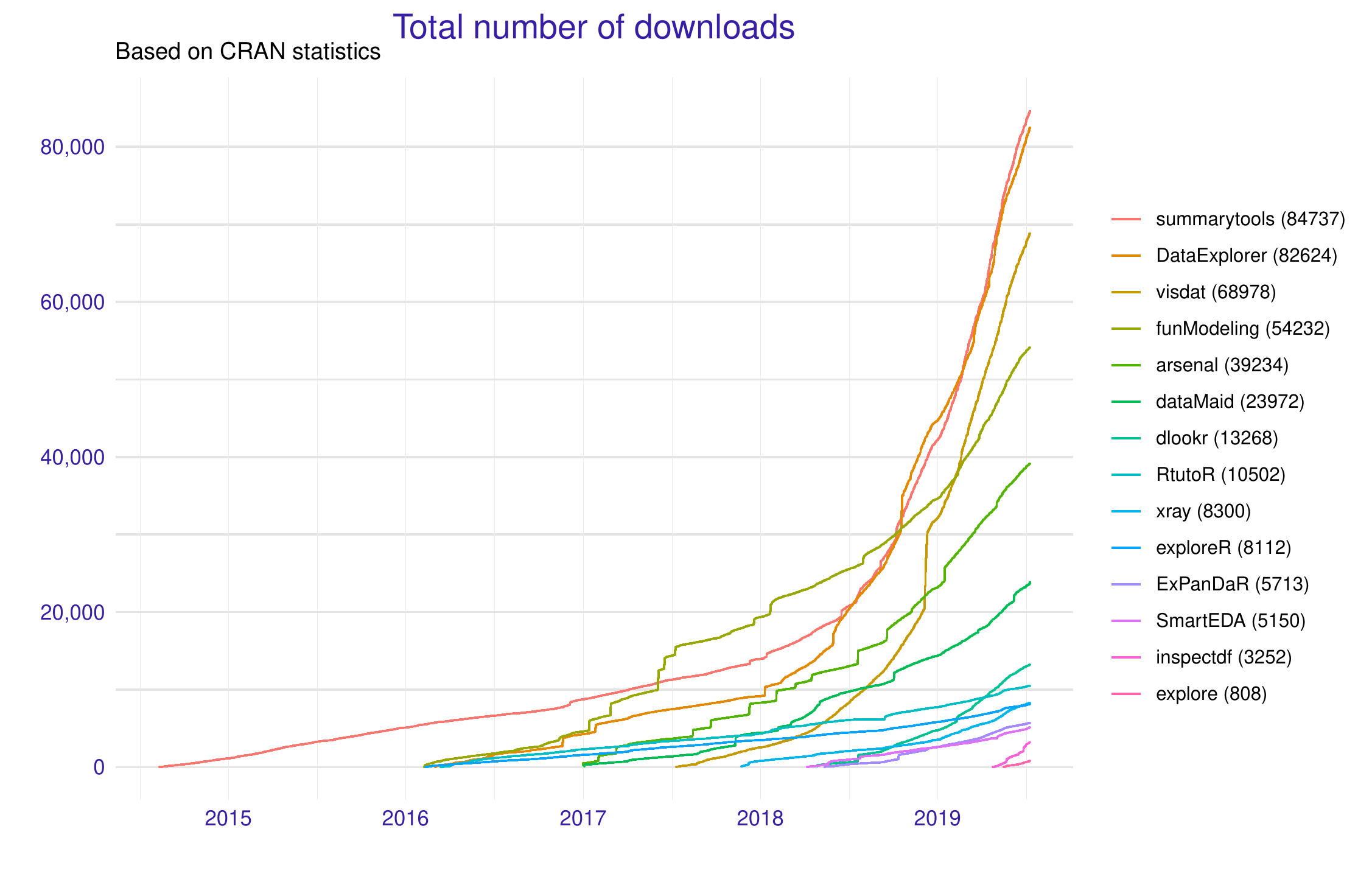}
    \caption{Trends in number of downloads of autoEDA packages available on CRAN since the first release. Data was gathered on 12.07.2019 with the help of the \CRANpkg{cranlogs} package \citep{cranlogs}.}
    \label{fig:downloads}
\end{figure}

Yet, model building is always preceded by a phase of understanding the problem, understanding of a domain and exploration of a data set.
Usually, in the process of the data analysis much more time is spent on data preparation and exploration than on model tuning. This is why the current bottleneck in data analysis is in the exploratory data analysis (EDA) phase.
Recently, a number of tools were developed to automate or speed up the part of the summarizing data and discovering patterns. 
Since the process of building predictive models automatically is referred to as autoML, we will dub the automation of data exploration autoEDA.
The surge in interest in autoEDA tools\footnote{Access the raw data with \code{archivist::aread("mstaniak/autoEDA-resources/autoEDA-paper/aec9")}} is evident in the Figure \ref{fig:downloads}.
Table \ref{table:popularity} describes the popularity of autoEDA tools measured as the number of downloads from CRAN and usage statistics from Github\footnote{Access the data with \code{archivist::aread("mstaniak/autoEDA-resources/autoEDA-paper/50a7")}}.

There is an abundance of R libraries that provide functions for both graphical and descriptive data exploration.
Here, we restrict our attention to packages that aim to automatize or significantly speed up the process of exploratory data analysis for tabular data.
Such tools usually work with full data frames, which are processed in an automatic or semi-automatic manner, for example by guessing data types, and return summary tables, groups of plots or full reports.
Currently, there is no CRAN Task View dedicated to packages for automated Exploratory Data Analysis and neither was there any repository that would catalog them\footnote{The first author maintains a list of papers related to autoEDA and software tools in different languages at \url{https://github.com/mstaniak/autoEDA-resources}}.
Here, we make a first attempt to comprehensively describe R tools for autoEDA. 
We chose two types of packages. 
The first group explicitly aims to automate EDA, as stated in the description of the package. 
These includes packages for \textit{fast}, \textit{easy}, \textit{interactive} or \textit{automated} data exploration.
The second group contains packages that create data summaries.
These packages were included, as long as they address at least two analysis goals listed in Table \ref{tab:tasks}.
We do not describe in detail packages that are either restricted to one area of application (for example \pkg{RBioPlot} \citep{rbioplot} package dedicated to biomolecular data or \pkg{intsvy} \citep{intsvy} package focused on international large-scale assessments), designed for one specific task (for example creating tables), or in an early development phase.
Some of the more task-specific packages are briefly discussed in Section \ref{other}.
Some packages, such as \CRANpkg{radiant} \citep{radiant} cover the full analysis pipeline and, as such, are too general for our purposes, even though they include an EDA module.

This paper has two main goals.
First is to characterize existing R packages for automated Exploratory Data Analysis and compare their ranges of capabilities.
To our best knowledge, this is first such a review.
Previously, a smaller comparison of seven packages was done in \cite{smartedapaper}. 
Second is to identify areas, where automated data exploration could be improved.
In particular, we are interested in gauging the potential of AI-assisted EDA tools.

The first goal is addressed in Sections \ref{chapter:packages} \textit{R packages for automated EDA} and \ref{chapter:feature} \textit{Feature comparison} where we first briefly describe each package and the compare, how are different EDA tasks are tackled by these packages.
Then, in Section \ref{chapter:summary} \textit{Summary}, we compile a list of strong and weak points of the automated EDA software and detail some open problems.

\begin{table}[ht]
\centering
\setlength\tabcolsep{4pt}
\begin{tabular}{l|rrr|rrrrr}
  \toprule
 & \multicolumn{3}{c|}{CRAN} & \multicolumn{5}{c}{GitHub} \\ 
package & downl. & debut & age & stars & commits & contrib. & issues & forks \\ 
arsenal & 39234 & 2016-12-30 & 2y 6m &  59 & 637 &   3 & 200 &   4 \\ 
autoEDA & - & - & - &  41 &  20 &   1 &   4 &  12 \\ 
DataExplorer & 82624 & 2016-03-01 & 3y 4m & 235 & 187 &   2 & 121 &  44 \\ 
dataMaid & 23972 & 2017-01-02 & 2y 6m &  68 & 473 &   2 &  45 &  18 \\ 
dlookr & 13268 & 2018-04-27 & 1y 2m &  35 &  54 &   3 &   9 &  12 \\ 
ExPanDaR & 5713 & 2018-05-11 & 1y 2m &  32 & 197 &   2 &   3 &  14 \\ 
explore & 808 & 2019-05-16 & 0y 1m &  15 & 114 &   1 &   1 &   0 \\ 
exploreR & 8112 & 2016-02-10 & 3y 5m &   1 &   1 &   1 &   0 &   0 \\ 
funModeling & 54232 & 2016-02-07 & 3y 5m &  58 & 126 &   2 &  13 &  18 \\ 
inspectdf & 3252 & 2019-04-24 & 0y 2m & 117 & 200 &   2 &  12 &  11 \\ 
RtutoR & 10502 & 2016-03-12 & 3y 3m &  13 &   7 &   1 &   4 &   8 \\ 
SmartEDA & 5150 & 2018-04-06 & 1y 3m &   4 &   4 &   1 &   1 &   2 \\ 
summarytools & 84737 & 2014-08-11 & 4y 11m & 255 & 981 &   6 &  76 &  33 \\ 
visdat & 68978 & 2017-07-11 & 2y 0m & 313 & 426 &  12 & 122 &  39 \\ 
xray & 8300 & 2017-11-22 & 1y 7m &  63 &  33 &   4 &  10 &   5 \\ 
  \bottomrule
\end{tabular}
\caption{Popularity of R packages for autoEDA among users and package developers. First two columns summarise CRAN statistics, last five columns summarise package development at GitHub. When a repository owned by the author is not available, the data were collected from a CRAN mirror repository. Data was gathered on 12.07.2019.}
\label{table:popularity}
\end{table}

\subsection{The tasks of Exploratory Data Analysis}\label{chapter:tasks}

Exploratory Data Analysis is listed as an important step in most methodologies for data analysis \citep{MDP, r4datascience}. One of the most popular methodologies, the CRISP-DM  \citep{crispdm}, lists the following phases of a data mining project:
\begin{enumerate}
    \item Business understanding.
    \item Data understanding.
    \item Data preparation.
    \item Modeling.
    \item Evaluation.
    \item Deployment.
\end{enumerate}
Automated EDA tools aim to make the Data understanding phase as fast and as easy as possible.
This part of a project can be further divided into smaller tasks.
These include a description of a dataset, data exploration, and data quality verification.
All these tasks can be achieved both by providing descriptive statistics and numerical summaries and by visual means.
AutoEDA packages provide functions to deal with these challenges.
Some of them are also concerned with simple variable transformations and data cleaning.
Both these tasks belong in the Data preparation phase, which precedes and supports the model building phase.
Let us notice that business understanding is affected by data understanding, which makes this part of the analysis especially important.

Goals of autoEDA tools are summarised in Table \ref{tab:tasks}.
The \textit{Phase} and \textit{Tasks} columns are taken from the CRISP-DM standard, while \textit{Type} and \textit{Examples} columns provide examples based on current functionalities of autoEDA packages.

\begin{table}[ht]
    \centering
    \begin{tabular}{c|c|c|c}
         Phase & Task & Type & Examples \\
        \toprule
        \multirow{9}{*}{Data understanding} & \multirow{3}{*}{Data description} & dimensions & variables number\\
        & & variables & variable type \\
        & & meta-data & size in RAM \\
        \cmidrule{2-4}
        & \multirow{3}{*}{Data validity} & invalid values & typos \\
        & & missing values & \texttt{NA} count \\
        & & atypical values & outliers \\
        \cmidrule{2-4}
        & \multirow{3}{*}{Data exploration} & univariate & histogram \\
        & & bivariate & scatter plot\\
        & & multivariate & Parallel coord. plot \\
        \midrule
         \multirow{6}{*}{Data preparation} & \multirow{2}{*}{Data cleaning} & Imputation & Impute mean \\
         & & Outlier treatment & Impute median \\
        \cmidrule{2-4}
         & \multirow{4}{*}{Derived attributes} & Dimension reduction & PCA \\
         & & \multirow{2}{*}{Continuous} & Box-Cox transform \\ 
         & & & Binning \\
         & & Categorical & Merge rare factors \\
         \bottomrule
    \end{tabular}
    \caption{Early phases of data mining project according to CRISP-DM standard, their specific goals and examples of how they are aided by autoEDA tools. \citep{crispdm}}
    \label{tab:tasks}
\end{table}

Each task should be summarised in a report, which makes reporting another relevant problem of autoEDA. 
Uni- and bivariate data exploration is a part of the analysis that is most thoroughly covered by the existing autoEDA tools. 
The form of univariate summaries depends on the variable type. For numerical variables, most packages provide descriptive statistics such as centrality and dispersion measures.
For categorical data, unique levels and associated counts are reported.
Bivariate relationships descriptions display either dependency between one variable of interest and all other variables, which includes contingency tables, scatter plots, survival curves, plots of distribution by values of a variable (histograms, bar plots, box plots), or between all pairs of variables (correlation matrices and plots), or chosen pairs of variables.

\section{R packages for automated EDA}\label{chapter:packages}

In this section, fifteen R libraries are shortly summarised.
One of them is only available on GitHub (\pkg{autoEDA}), all others are available at CRAN.
For each library, we include example outputs.
The exact versions of packages that were used to create them can be found in the reference section.
All examples are based on a subset of \code{typical\_data}\footnote{Access the data with \code{archivist::aread("mstaniak/autoEDA-resources/autoEDA-paper/278c7")}} dataset from \CRANpkg{visdat} package.
Whenever possible, \CRANpkg{archivist} \citep{archivist} hooks are provided for easy access to the presented objects.
When a function call only gives side-effects, a link is provided to the full result (PDF/PNG files).
Tables were prepared with the \CRANpkg{xtable} package \citep{xtable}.

\subsection{The \pkg{arsenal} package}

\begin{figure}
    \centering
    \includegraphics[scale=0.85]{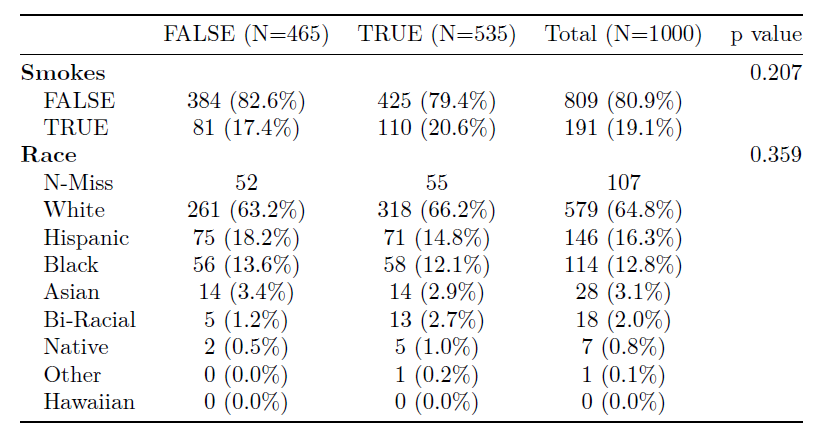}
    \caption{An example output from the \code{arsenal::tableby} function saved using \code{arsenal::write2} (\pkg{arsenal} v 2.0). \texttt{Smokes} and \texttt{Race} variables are compared by the levels of \texttt{Died} variable.}
    \label{fig:arsenaltable}
\end{figure}

The \CRANpkg{arsenal} package \citep{arsenal} is a set of four tools for data exploration:
\begin{enumerate}
    \item table of descriptive statistics and p-values of associated statistical tests, grouped by levels of a target variable (the so-called \textit{Table 1}). Such a table can also be created for paired observation, for example longitudinal data (\code{tableby} and \code{paired} functions),
    \item comparison of two data frames that can detect shared variables (\code{compare} function),
    \item frequency tables for categorical variables (\code{freqlist} function),
    \item fitting and summarizing simple statistical models (linear regression, Cox model, etc) in tables of estimates, confidence intervals and p-values (\code{modelsum} function).
\end{enumerate}
Results of each function can be saved to a short report using the \code{write2} function.
An example\footnote{Access the table with \code{archivist::aread("mstaniak/autoEDA-resources/autoEDA-paper/d951")}} can be found in Figure \ref{fig:arsenaltable}.

A separate vignette is available for each of the functions. 
\code{arsenal} is the most statistically-oriented package among reviewed libraries.
It borrows heavily from SAS-style procedures used by the authors at the Mayo Clinic.

\subsection{The \pkg{autoEDA} package}

\pkg{autoEDA} package \citep{autoeda}  is a GitHub-based tool for univariate and bivariate visualizations and summaries.
The \code{dataOverview} function returns a data frame that describes each feature by its type, number of missing values, outliers and typical descriptive statistics. 
Values proposed for imputation are also included.
Two outlier detection methods are available: Tukey and percentile-based.
A PDF report can be created using the \code{autoEDA} function.
It consists of the plots of distributions of predictors grouped by outcome variable or distribution of outcome by predictors.

\begin{figure}
    \centering
    \includegraphics[width=5in,height=3.5in]{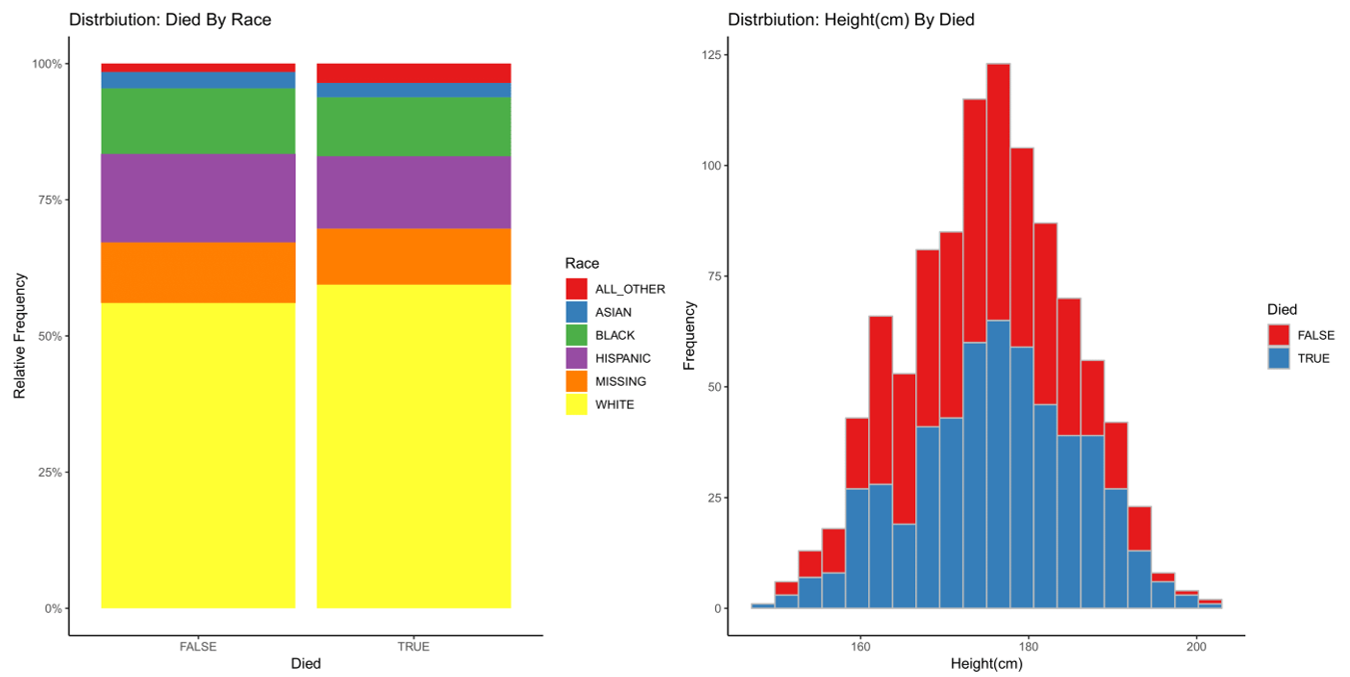}
    \caption{Sample pages from the report generated by the \code{autoEDA::autoEDA} function (\pkg{autoEDA} v. 1.0) displaying bivariate relationships between the target and explanatory variable.}
    \label{fig:autoeedaex}
\end{figure}

The package can be found on Xander Horn's GitHub page: \url{https://github.com/XanderHorn/autoEDA}.
It does not include a vignette, but a short introductory article was published to LinkedIn \citep{autoedablog} and similar examples can be found in the readme of the project.
Plots from a report\footnote{Find the full report at \url{https://github.com/mstaniak/autoEDA-resources/blob/master/autoEDA-paper/plots/autoEDA/autoEDA_report.pdf}} generated by \pkg{autoEDA} are displayed in Figure \ref{fig:autoeedaex}.

\subsection{The \pkg{DataExplorer} package}

\begin{figure}
    \centering
    \includegraphics[width=\textwidth]{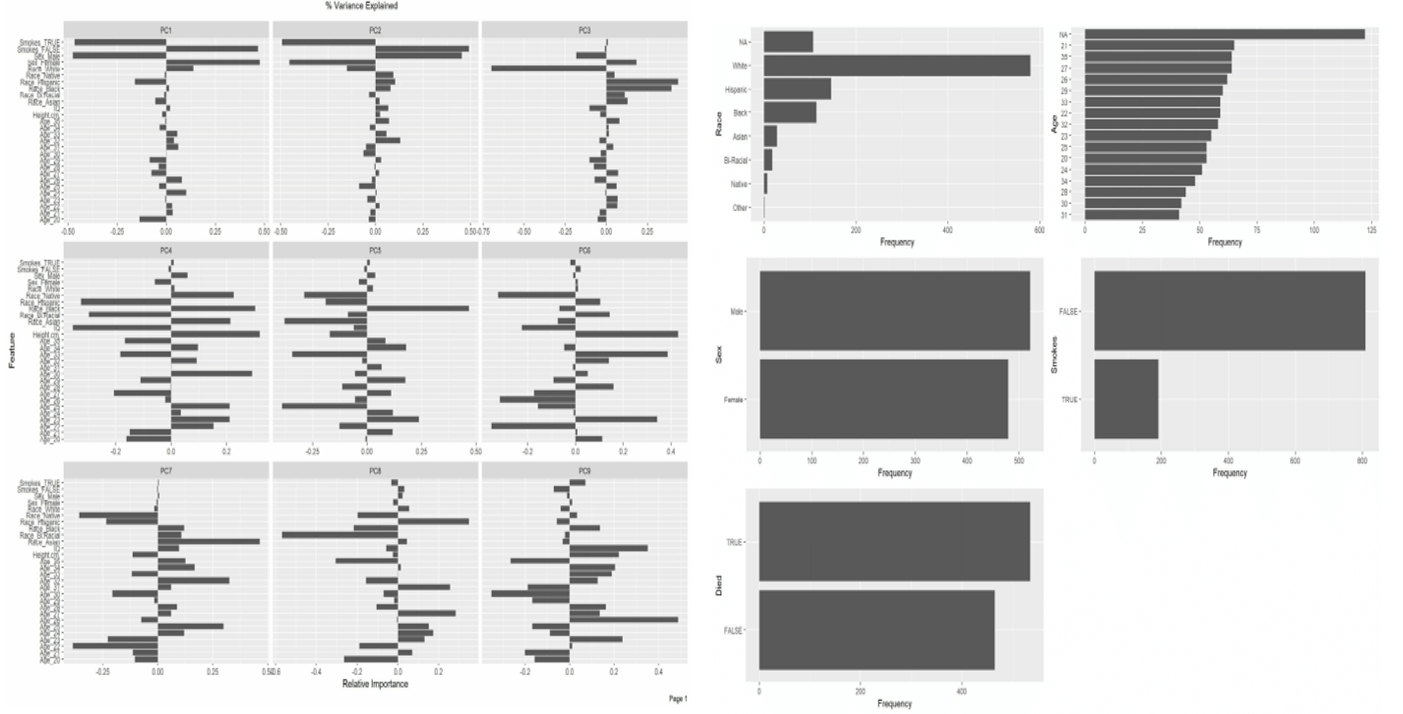}
    \caption{A visualization of PCA results - correlation with original variables for each principal component - and a \textit{wall of bar plots} taken from a report generated by the \code{DataExplorer::create\_report} function (\pkg{DataExplorer} v. 0.7).}
    \label{fig:dataexplorerreport}
\end{figure}

\CRANpkg{DataExplorer} \citep{dataexplorer} is a recent package that helps automatize EDA and simple data transformations.
It provides functions for:
\begin{enumerate}
    \item whole dataset summary: dimensions, types of variables, missing values, etc (\code{introduce} and \code{plot\_intro} functions),
    \item missing values profile as a plot of missing values fraction per column (\code{plot\_missing} function) or summary statistics and suggested actions (\code{profile\_missing} function),
    \item plotting distributions of variables, separately numerical and categorical (\code{plot\_histogram} and \code{plot\_bar} functions),
    \item QQ Plots (\code{plot\_qq} function),
    \item plotting correlation matrices (\code{plot\_correlation} function),
    \item visualizing PCA results by plotting percentage of explained variance and correlations with each original feature for every principal component (\code{plot\_prcomp} function),
    \item plotting relationships between the target variable and predictors  - scatterplots and boxplots (\code{plot\_scatterplot} and \code{plot\_boxplot} functions),
    \item data transformation: replacing missing values by a constant (\code{set\_missing} function), grouping sparse categories (\code{group\_category} function), creating dummy variables, dropping columns (\code{dummify}, \code{drop\_features} functions) and modifying columns (\code{update\_columns} function).
\end{enumerate}
The \code{create\_report} function generates a report.
By default, it consists of all the above points except for data transformations and it can be further customized.
An introductory vignette \textit{Introduction to DataExplorer} that showcases all the functionalities is included in the package.
It is noticeable that the package almost entirely relies on visual techniques.
Plots taken from an example report\footnote{Access the full report \url{https://github.com/mstaniak/autoEDA-resources/blob/master/autoEDA-paper/plots/DataExplorer/dataexplorer_example.pdf}} are presented in Figure \ref{fig:dataexplorerreport}.

\subsection{The \pkg{dataMaid} package} 

The \CRANpkg{dataMaid} \citep{datamaid} package has two central functions: the \code{check} function, which performs checks of data consistency and validity, and \code{summarize}, which summarizes each column.
Another function, \code{makeDataReport}, automatically creates a report in PDF, DOCX or HTML format. 
The goal is to detect missing and unusual - outlying or incorrectly encoded - values.
The report contains a whole dataset summary: variables and their types, number of missing values, and univariate summaries in the form of descriptive statistics, histograms/bar plots and an indication of possible problems.

\begin{figure}
    \centering
    \includegraphics[width=5in,height=3.5in]{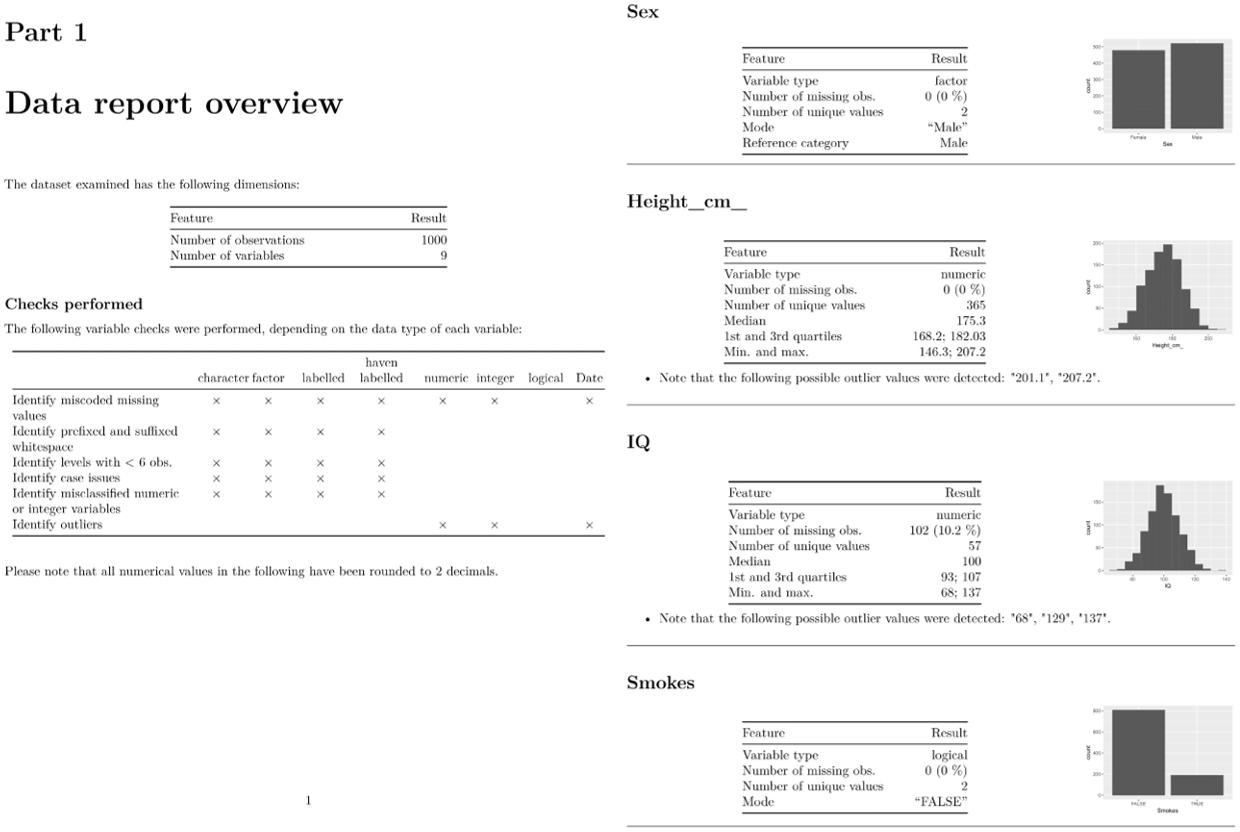}
    \caption{Two pages from a data validity report generated using the \code{dataMaid::makeDataReport} function (\pkg{dataMaid} v. 1.2). Atypical values are listed under the variable summary.}
    \label{fig:datamaidexample}
\end{figure}

User-defined checks and summaries can be also included in the analysis.
The vignette \textit{Extending dataMaid} explains how to define them.
It is also possible to customize the report.
In particular, variables for which no problems were detected can be omitted.
An example report\footnote{Find the full report at  \url{https://github.com/mstaniak/autoEDA-resources/blob/master/autoEDA-paper/plots/dataMaid/dataMaid\_report.pdf}} can be found in Figure \ref{fig:datamaidexample}. 

\subsection{The \pkg{dlookr} package}

The \CRANpkg{dlookr} \citep{dlookr} package provides tools for 3 types of analysis: data diagnosis including correctness, missing values, outlier detection; exploratory data analysis; and variable transformations: imputation, dichotomization, and transformation of continuous features.
It can also automatically generate a PDF report for all these analyses.

For data diagnosis, types of variables are reported along with counts of missing values and unique values.
Variables with a low proportion of unique values are described separately.
All the typical descriptive statistics are provided for each variable.
Outliers are detected and distributions of variables before and after outlier removal are plotted.
Both missing values and outliers can be treated using \code{impute\_na} and \code{impute\_outlier} functions.

In the EDA report, descriptive statistics are presented along with normality tests, histograms of variables and their transformations that reduce skewness: logarithm and root square.
Correlation plots are shown for numerical variables. 
If the target variable is specified, plots that show the relationship between the target and each predictor are also included.

A transformation report compares descriptive statistics and plots for each variable before and after imputation, skewness-removing transformation and binning.
If the right transformation is found among the candidate transformations, it can be applied to the feature through one of the \code{binning}, \code{binning\_by}, or \code{transform} functions.

\begin{figure}
    \centering
    \includegraphics[scale=0.7]{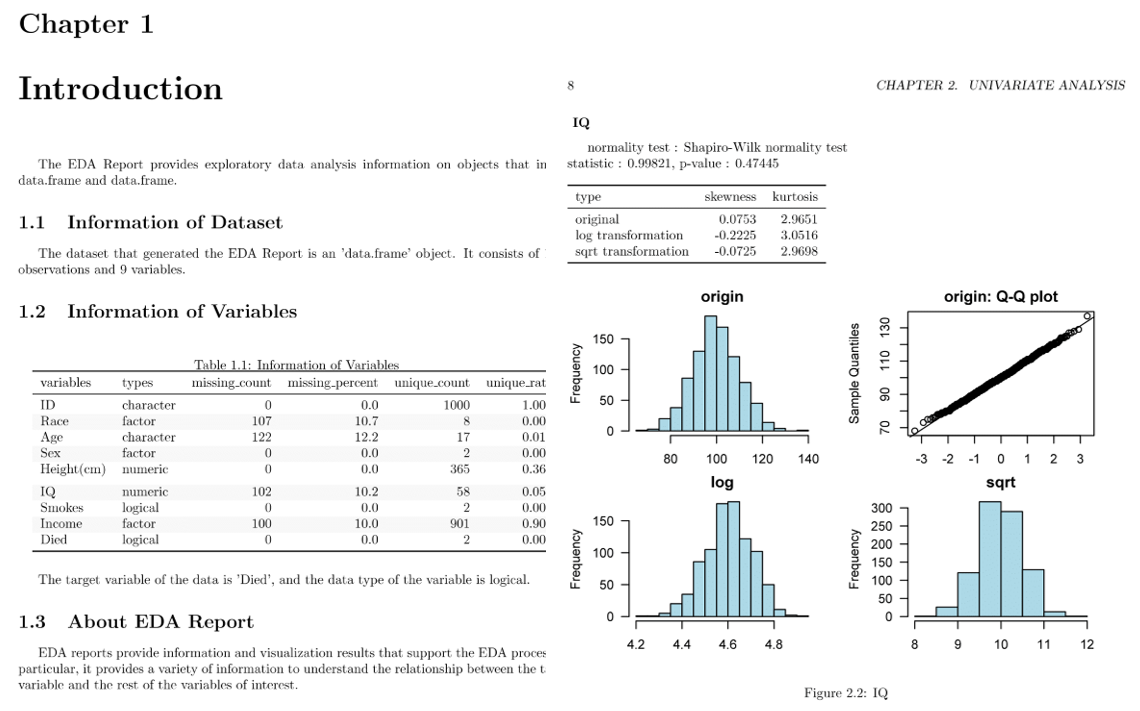}
    \caption{Two pages from a report generated by the \code{dlookr::eda\_report} function (\pkg{dlookr} v. 0.3.8). First, the dataset is summarised, than each variable is described. Optionally, plots of bivariate relationships can be added.}
    \label{fig:dlookreda}
\end{figure}

Every operation or summary presented in the reports can also be performed manually.
A dedicated vignette explains each of the main functionalities (\textit{Data quality diagnosis}, \textit{Data Transformation}, \textit{Exploratory Data Analysis} vignettes).
An example\footnote{Access the full report at \url{https://github.com/mstaniak/autoEDA-resources/blob/master/autoEDA-paper/plots/dlookr/dlookr_eda.pdf}} taken from one of the reports can be found in Figure \ref{fig:dlookreda}.

\subsection{The \pkg{ExPanDaR} package}

Notably, while the \CRANpkg{ExPanDaR} package \citep{panel} was designed for panel data exploration, it can also be used for standard EDA after adding an artificial constant time index.
In this case, the package offers interactive \pkg{shiny} application for exploration.
Several types of analysis are covered: 
\begin{enumerate}
    \item missing values and outlier treatment,
    \item univariate summaries (descriptive statistics) and plots (histograms/bar plots),
    \item bivariate analysis via correlation matrices and plots. Interestingly, scatter plots can be enriched by associating size and color of points with variables,
    \item multivariate regression analysis.
\end{enumerate}
For each functionality of the application, there is a corresponding standalone function.

\begin{figure}
    \centering
    \includegraphics[scale=0.6]{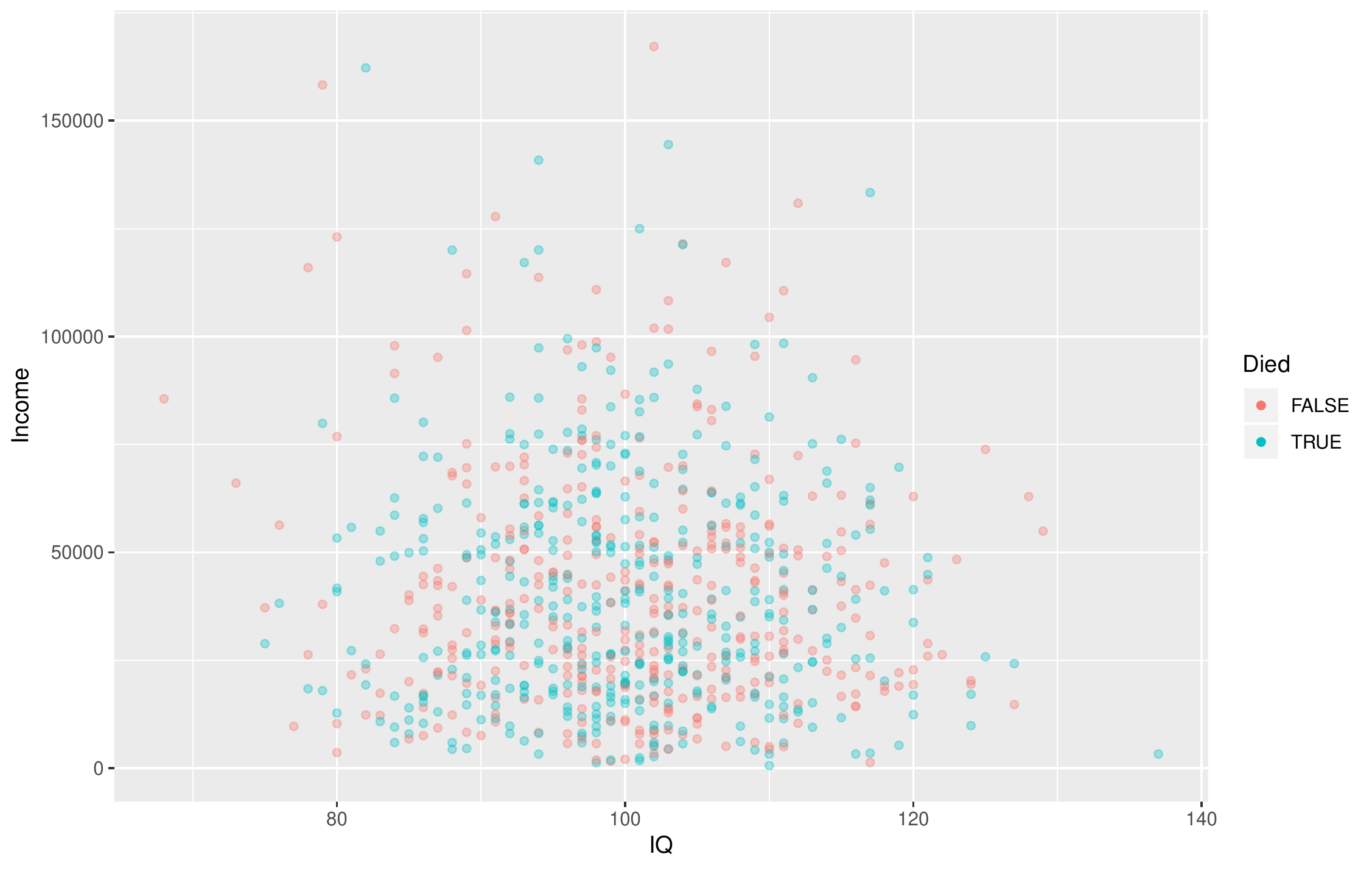}
    \caption{Scatter plot with of \code{Income} and \code{IQ} variables with \code{Died} variable denoted by the color. Created with the \code{prepare\_scatter\_plot} function (\pkg{ExPanDaR} v0.4.0).}
    \label{fig:expandar}
\end{figure}

Three vignettes describe how the library can be used for data exploration (\textit{Using the functions of the ExPanDaR package}), how to customize it (\textit{Customize ExPanD}) and how to analyze panel data (\textit{ Using ExPanD for Panel Data Exploration})
Example instances of \pkg{ExPanDaR} \pkg{shiny} applications are available online. 
Links and other examples can be found in the GitHub repository of the package: \url{https://github.com/joachim-gassen/ExPanDaR}.
An example of a scatter plot\footnote{Access the R object with the \code{archivist::aread("mstaniak/autoEDA-resources/autoEDA-paper/9c5d")}} created by the package can be found in Figure \ref{fig:expandar}.

\subsection{The \pkg{explore} package}

\begin{figure}
    \centering
    \includegraphics[scale=0.5]{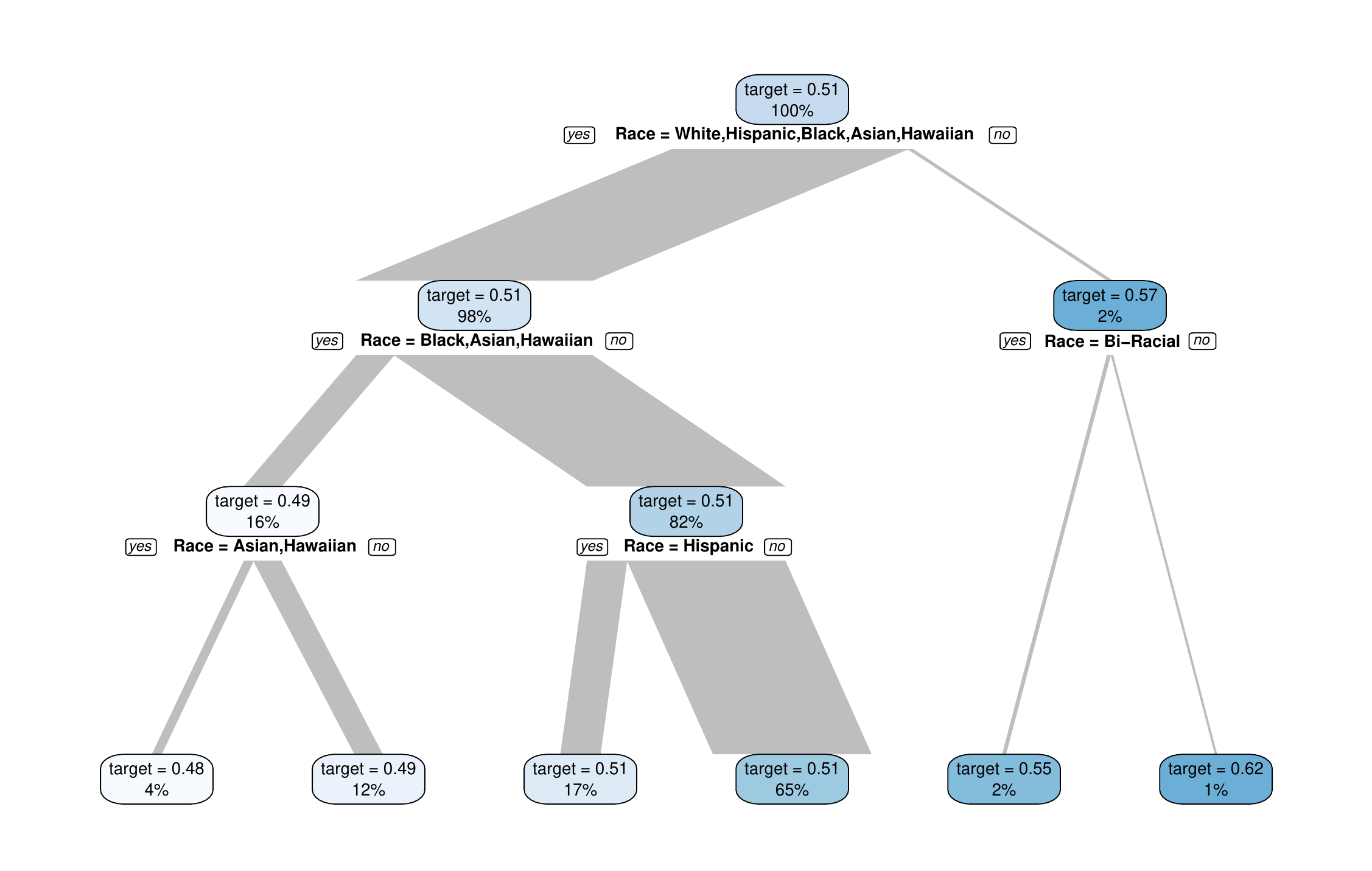}
    \caption{A decision tree fitted using the \code{explain\_tree} function (\pkg{explore} v. 0.4.3). The tree can also be based on multiple explanatory variables.}
    \label{fig:decisiontree}
\end{figure}

The functionalities of the \CRANpkg{explore} package \citep{explore} can be accessed in three ways: through an interactive \CRANpkg{shiny} \citep{shiny} application, through an automatically generated HTML report or via standalone functions.
In addition to data exploration, relationships with a binary target can be explored.
The package includes functions for 
\begin{enumerate}
    \item full dataset summaries - dimensions, data types, missing values and summary statistics (\code{describe} function),
    \item uni- and bivariate visualizations, including density plots, bar plots and boxplots (a family of \code{explore} functions, in particular \code{explore\_all} function that creates plots for all variables),
    \item modeling based on decision trees or logistic regression (\code{explain\_tree} and \code{explain\_logreg} functions).
\end{enumerate}
All result can be saved to HTML via the \code{report} function.
Dataset and variable summaries can also be save to an MD file using the \code{data\_dict\_md} function\footnote{Find examples at \url{https://github.com/mstaniak/autoEDA-resources/tree/master/autoEDA-paper/plots/explore}}.
The \textit{explore} vignette includes a thorough description of the package.
An example decision tree\footnote{Access the R object with the \code{archivist::aread("mstaniak/autoEDA-resources/autoEDA-paper/dc47")}.} can be found in Figure \ref{fig:decisiontree}.

\subsection{The \pkg{exploreR} package}

\begin{figure}
    \centering
    \includegraphics[scale=0.8]{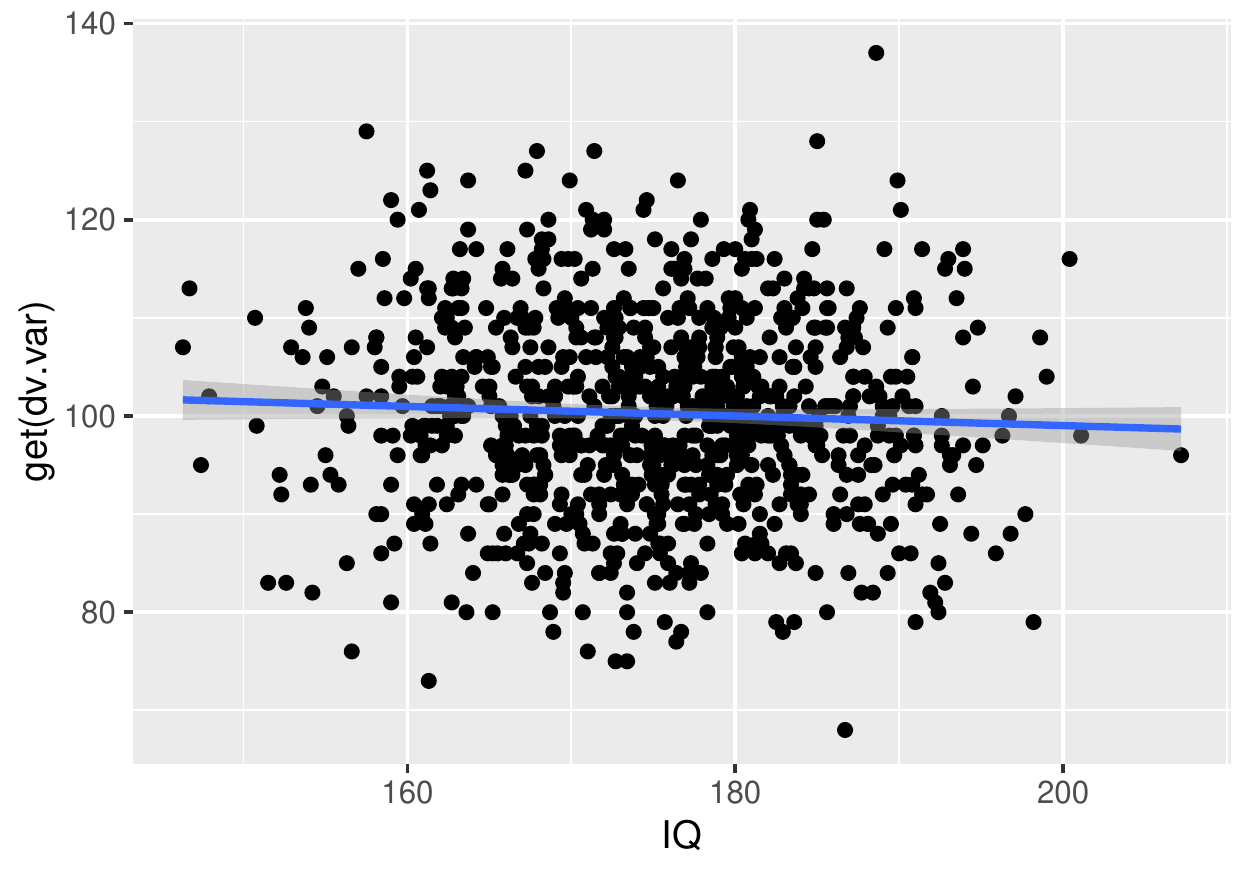}
    \caption{Univariate regression plot created using the \code{exploreR::massregplot} (\pkg{exploreR} v. 0.1).}
    \label{fig:explorer}
\end{figure}

The \CRANpkg{exploreR} package \citep{explorer} takes a unique approach to data exploration compared to other packages.
The analysis is based on linear regression.
There are three functionalities:
\begin{enumerate}
    \item fitting univariate regression model for each independent variable and summarizing the results in a table that consists of estimated parameters, p-values, and $R^{2}$ values (\code{masslm} function),
    \item plotting target variable against each independent variable along with the fitted least-squares line  (\code{massregplot} function),
    \item feature standardization by scaling to the interval $[0, 1]$ or subtracting mean and dividing by standard deviation.
\end{enumerate}
Regression plots can be saved to a PDF file.
A vignette called \textit{The How and Why of Simple Tools} explains all the functions and provides examples.
One of the regression plots\footnote{A PDF file with all the plots can be found at \url{https://github.com/mstaniak/autoEDA-resources/blob/master/autoEDA-paper/plots/exploreR.pdf}} is presented in Figure \ref{fig:explorer}.

\subsection{The \pkg{funModeling} package}

\begin{figure}
    \centering
    \includegraphics[scale=0.8]{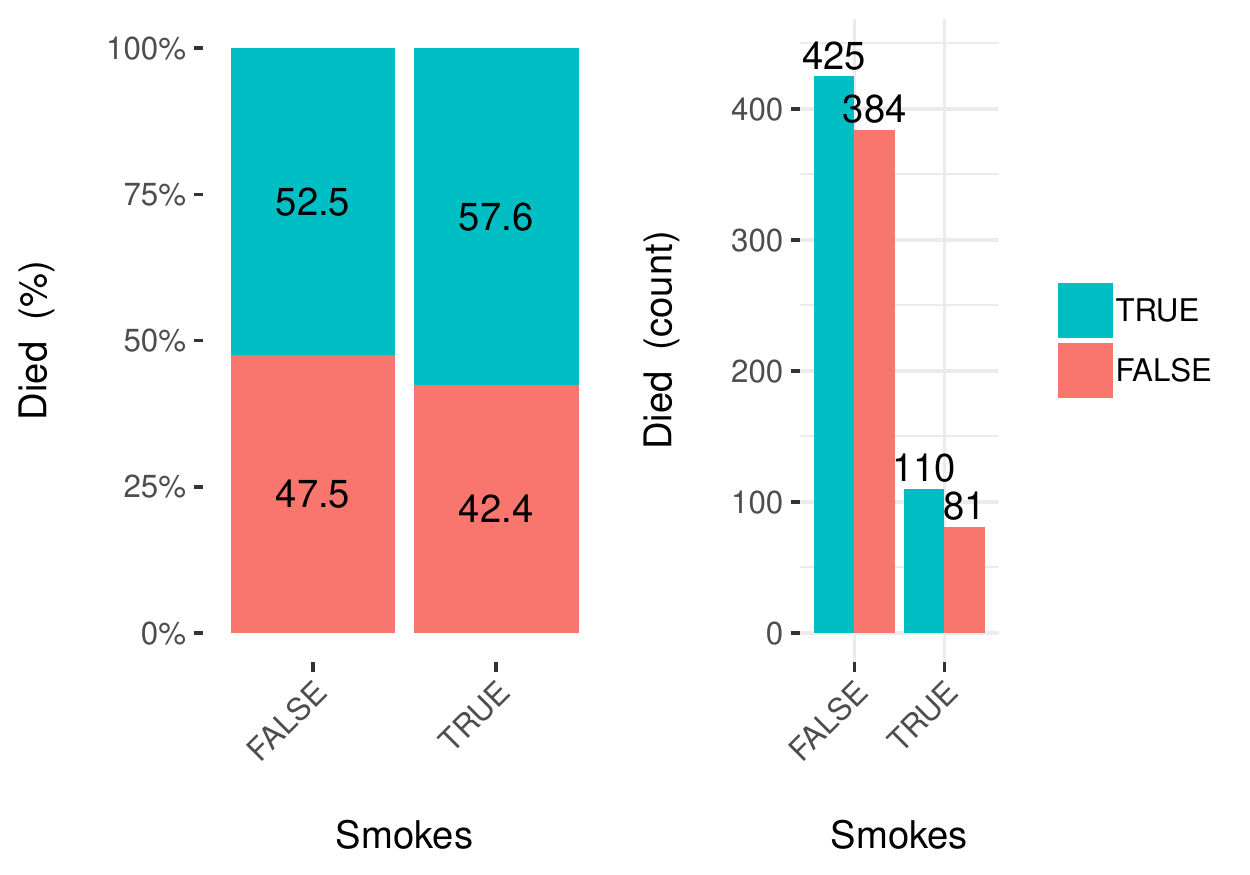}
    \caption{An example output from the \code{funModeling::cross\_plot} function (\pkg{funModeling} v. 1.7). Such a plot is drawn for every variable in the dataset or for a specified subset of variables. Continuous features are discretized.}
    \label{fig:funmodeling}
\end{figure}

The package \CRANpkg{funModeling} \citep{funmodeling} is a rich set of tools for EDA connected to the book \cite{datasciencelive}.
These tools include 
\begin{enumerate}
    \item dataset summary (\code{df\_status} function), 
    \item plots and descriptive statistics for categorical and numerical variables (\code{plot\_num}, \code{profiling\_num} and \code{freq} functions), 
    \item classical and information theory-based correlation analysis for target variable vs other variables - (\code{correlation\_table} function for numerical predictors, \code{var\_rank\_info} function for all predictors),
    \item plots of distribution of target variables vs predictors (bar plots, box plots and histograms via \code{cross\_plot} and \code{plotar} functions),
    \item quantitative analysis for binary target variables (\code{categ\_analysis} function),
    \item different methods of binning continuous features (\code{discretize\_df}, \code{convert\_df\_to\_categoric} and \code{discretize\_rgr} functions),
    \item variable normalization by transforming to the $[0, 1]$ interval (\code{range01} function),
    \item outlier treatment (\code{prep\_outliers}, \code{tukey\_outlier} and \code{hampel\_outlier} functions),
    \item gain and lift curves (\code{gain\_lift} function).
\end{enumerate}
It is the only library that encompasses visualizations related to predictive models and non-standard correlation analysis.
The range of tools covered by \pkg{funModeling} is very wide.
The package includes an exhaustive introduction vignette called \textit{\pkg{funModeling} quick-start}.
One of the bivariate visualizations\footnote{Find all the plots at \url{https://github.com/mstaniak/autoEDA-resources/tree/master/autoEDA-paper/plots/funmodeling}} offered by the package can be found in Figure \ref{fig:funmodeling}.

\subsection{The \pkg{inspectdf} package}

\begin{figure}
    \centering
    \includegraphics[scale=0.6]{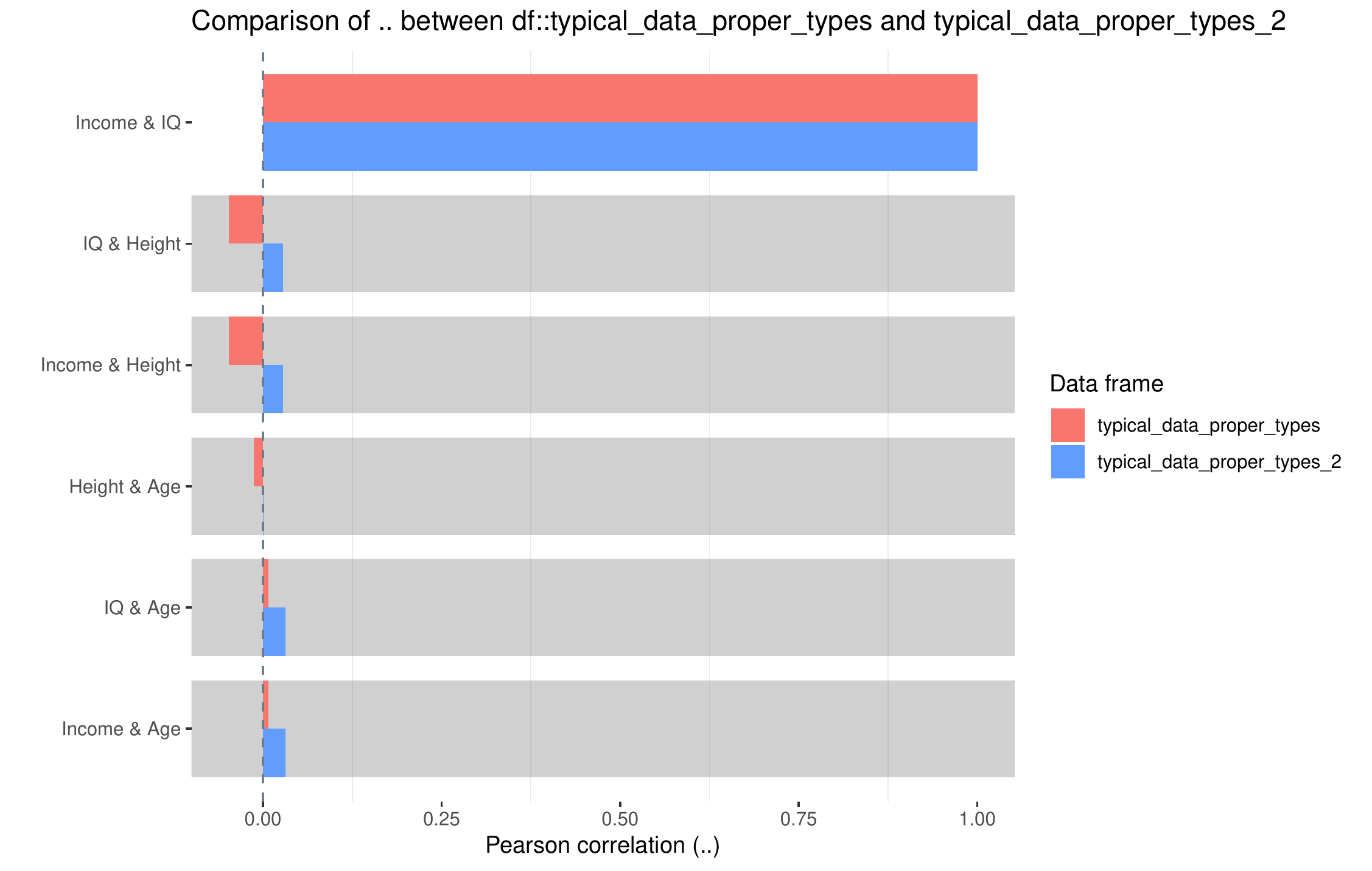}
    \caption{A comparison of correlations between numerical variables in two data frames. Plot created using the \pkg{inspectdf} package v. 0.0.3.}
    \label{fig:inspectdf}
\end{figure}

The \CRANpkg{inspectdf} package \citep{inspectdf} provides several tools for basic data exploration with a consistent interface. 
Each of the \code{inspect\_*} functions returns a data frame with summaries (and additional attributes).
The results can be then plotted using the \code{show\_plot} function.
The functions are related to three aspects of EDA:
\begin{enumerate}
    \item whole dataset can be summarised by numbers of missing values, number of variables of each type and memory used by each variable (\code{inspect\_na}, \code{inspect\_types} and \code{inspect\_mem} functions),
    \item univariate analysis is done via summary statistics and histograms for numerical variables (\code{inspect\_num} function), bar plots for categorical variables (\code{inspect\_cat} function). Additionally, factors dominated by a single level can be found with the \code{inspect\_imb} function,
    \item bivariate relationships are described by Pearson correlation coefficient for numerical variables (\code{inspect\_cor} function).
\end{enumerate}
Notably, each function can take two data frames as parameters and return their comparison.
An example of a correlation analysis plot comparing two data frames can be found in Figure \ref{fig:inspectdf}
While the library does not include a vignette, extensive documentation with examples is provided on the GitHub webpage of the project: \url{https://github.com/alastairrushworth/inspectdf}.

\subsection{The \pkg{RtutoR} package}

The \CRANpkg{RtutoR} package \citep{rtutor} is a tool for automated reporting. 
There are three options for creating a report that contains univariate and bivariate data summaries:
\begin{enumerate}
    \item plots can be created interactively in a \pkg{shiny} app (\code{launch\_plotter} function),
    \item the whole report can be generated from a \pkg{shiny} app that allows the user to tweak the report (\code{gen\_exploratory\_report\_app} function),
    \item the report can be created by a direct call to the \code{generate\_exploratory\_analysis\_ppt} function.
\end{enumerate}
The report is saved in the \texttt{PPTX} format.
Notably, this package can identify the top \texttt{k} relevant variables based on a chosen criterion, for example, information gain, and display plots only for these variables.
An example report can be found in the GitHub repository of the package\footnote{Find the report at \url{https://github.com/anup50695/RtutoR/blob/master/titanic_exp_report_2.pptx}}.
The package was introduced in an R-Bloggers blog post \citep{rtutorblog}.

\subsection{The \pkg{SmartEDA} package}

\begin{figure}
    \centering
    \includegraphics[width=5in,height=3.5in]{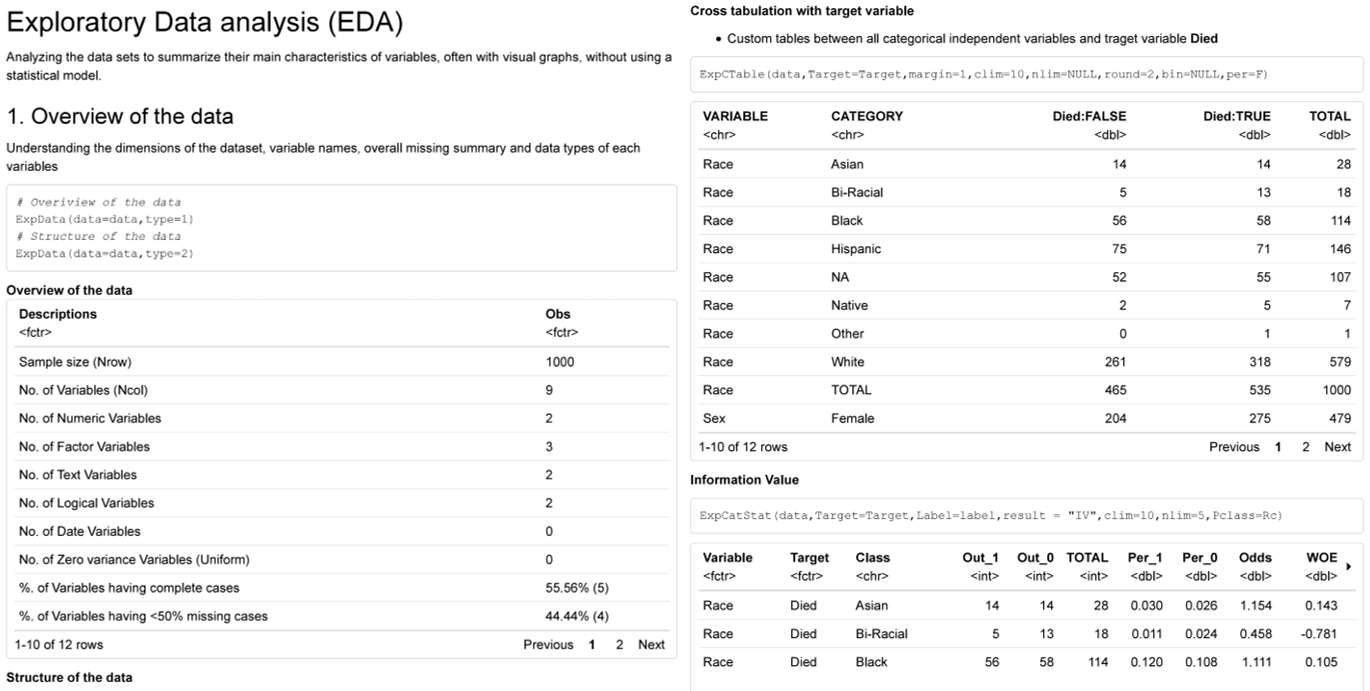}
    \caption{Sample pages from a report generated by the \code{SmartEDA::ExpReport} function (\pkg{SmartEDA} v. 0.3), including dataset overview and bivariate dependency for categorical variables.}
    \label{fig:smartedareport}
\end{figure}

The \CRANpkg{SmartEDA} package \citep{smarteda}, is focused entirely on data exploration through graphics and descriptive statistics. It does not provide any functions which modify existing variables.
The range of tools it includes is wide:
\begin{enumerate}
    \item dataset summary (\code{ExpData} function),
    \item descriptive statistics that may include correlation with target variable and density or bar plots (\code{ExpNumStat}, \code{ExpNumViz}, \code{ExpCatStat} and \code{ExpCatViz} functions). All visualizations may include the target variable,
    \item QQ plots (\code{ExpOutQQ} function),
    \item contingency tables (\code{ExpCTable} function),
    \item information value and Weight of the Evidence coding (\code{ExpWoETable}, \code{ExpInfoValue} functions),
    \item parallel coordinate plot for multivariate visualization (\code{ExpParcoord} function).
\end{enumerate}
Plotting functions return grids of \pkg{ggplot2} object.
The results can be written to a HTML report (\code{ExpReport} function).
There are also additional functionalities dedicated to \code{data.table} objects from \CRANpkg{data.table} package \citep{datatable}.
An introductory vignette called \textit{Explore data using SmartEDA (Intro)} is attached to the library.
Another vignette \textit{Custom summary statistics} describe customizing output tables.
The package is also described in the \cite{smartedapaper} paper.
Examples\footnote{A full report is available at \url{https://github.com/mstaniak/autoEDA-resources/blob/master/autoEDA-paper/plots/SmartEDA/smarteda_report_target.pdf}} can be found in Figure \ref{fig:smartedareport}.

\subsection{The \pkg{summarytools} package}

The \CRANpkg{summarytools} package \citep{summarytools} builds summary tables for whole datasets, individual variables, or pairs of variables. 
In addition, the output can be formatted to be included in \CRANpkg{knitr}\citep{knitr} or plain documents, HTML files and \CRANpkg{shiny} apps \citep{shiny}.
There are four main functionalities:
\begin{enumerate}
    \item whole dataset summary including variable types and a limited number of descriptive statistics, counts of unique values and missing values and univariate plots within the output table (\code{dfSummary} function),
    \item descriptive statistics, including skewness and kurtosis, for numerical variables, possibly grouped by levels of a factor (\code{descr}, \code{stby} functions),
    \item counts and proportions for levels of categorical features (\code{freq} function),
    \item contingency tables for pairs of categorical variables (\code{ctable} function).
\end{enumerate}
All results can be saved and displayed in different formats.
The package includes a vignette titled \textit{Introduction to summarytools}.
An example of univariate summaries\footnote{Access the R object with \code{archivist::aread("mstaniak/autoEDA-resources/autoEDA-paper/9e12")}.} can be found in Figure \ref{table:summarytoolstbl}.

\begin{table}[ht]
\centering
\begin{tabular}{rrr}
  \hline
 & Height(cm) & IQ \\ 
  \hline
Mean & 175.09 & 100.23 \\ 
  Std.Dev. & 9.83 & 10.03 \\ 
  Min & 146.30 & 68.00 \\ 
  Q1 & 168.20 & 93.00 \\ 
  Median & 175.30 & 100.00 \\ 
  Q3 & 182.05 & 107.00 \\ 
  Max & 207.20 & 137.00 \\ 
  MAD & 10.38 & 10.38 \\ 
  IQR & 13.83 & 14.00 \\ 
  CV & 0.06 & 0.10 \\ 
  Skewness & -0.08 & 0.08 \\ 
  SE.Skewness & 0.08 & 0.08 \\ 
  Kurtosis & -0.30 & -0.04 \\ 
  N.Valid & 1000.00 & 898.00 \\ 
  \% Valid & 100.00 & 89.80 \\ 
   \hline
\end{tabular}
\caption{An example table of descriptive statistics generated by the \code{summarytools::descr} function (\pkg{summarytools} v. 0.9.2).}
\label{table:summarytoolstbl}
\end{table}

\subsection{The \pkg{visdat} package}

\begin{figure}
    \centering
    \includegraphics{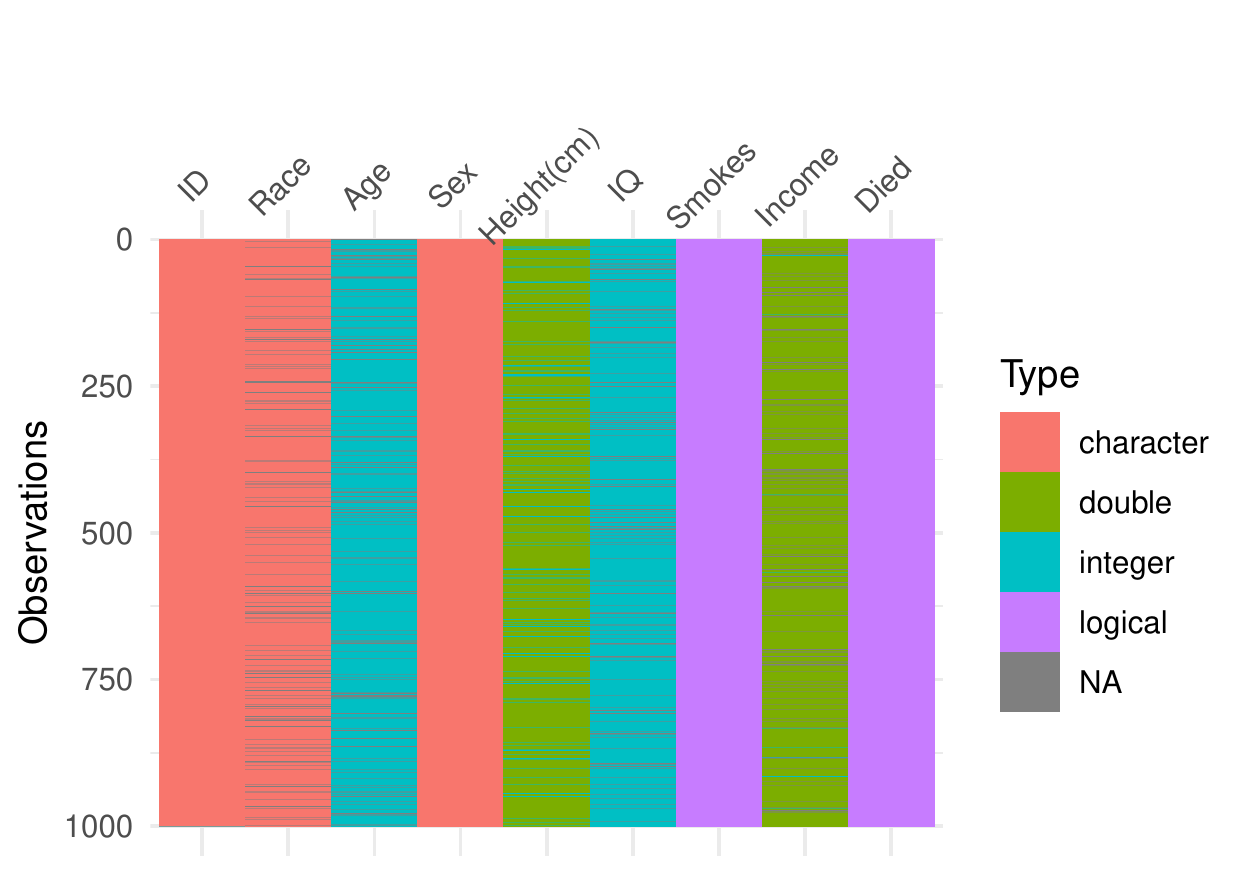}
    \caption{Example output of the \code{visdat::vis\_guess} function (\pkg{visdat} v. 0.5.3), which displays types of each value in the data frame and the missing values. We can see that the \texttt{Age} variable consists of integer values, even though it is coded as a \texttt{character}.}
    \label{fig:visguess}
\end{figure}

The package \pkg{visdat} \citep{visdat} is maintained by rOpenSci. It consists of six functions that help visualize:
\begin{enumerate}
    \item variables types and missing data (\code{vis\_dat} function), 
    \item types of each value in each column (\code{vis\_guess} function),
    \item clusters of missing values (\code{vis\_miss} function),
    \item differences between the two datasets (\code{vis\_compare} function),
    \item where given conditions are satisfied in the data (\code{vis\_expect} function),
    \item correlation matrix for the numerical variables (\code{vis\_cor} function).
\end{enumerate}
Each of these functions returns a single \CRANpkg{ggplot2} \citep{ggplot} plot that shows a rectangular representation of the dataset where the expected information is denoted by colors. 
An example of this visualization\footnote{Access the plot object with \code{archivist::aread("mstaniak/autoEDA-resources/autoEDA-paper/3cfd")}} can be seen in Figure \ref{fig:visguess}.

The package includes a vignette \textit{Using visdat} that provides examples for all package options. 
Interestingly, it is the only package that uses solely visual means of exploring the data.

\subsection{The \pkg{xray} package}

\begin{figure}
    \centering
    \includegraphics[scale=0.7]{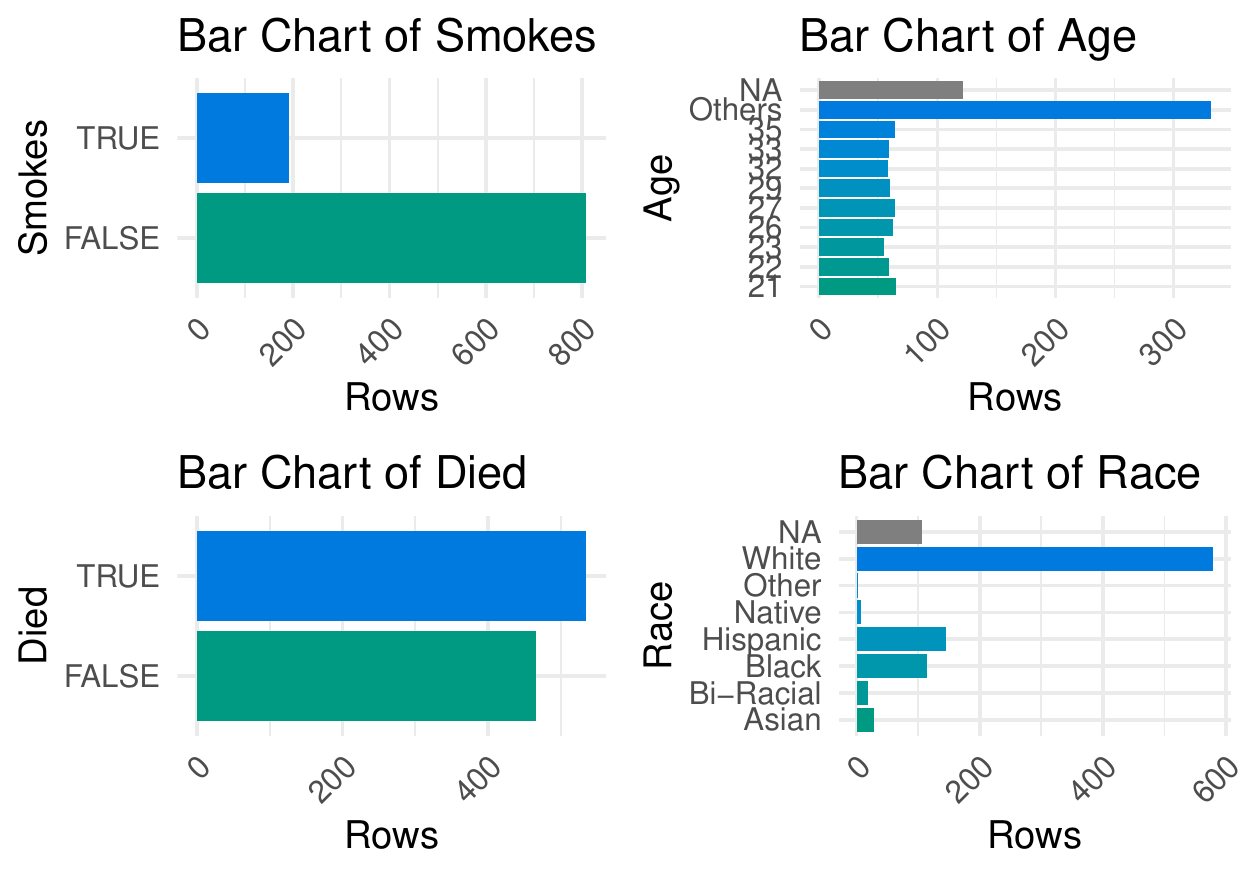}
    \caption{Example output from the \code{xray::distributions} function (\pkg{xray} v. 0.2). Such plots are created for each variable in the dataset along with a table of descriptive statistics.}
    \label{fig:xrayplot}
\end{figure}

The \CRANpkg{xray} \citep{xray} package has three functions for the analysis of data prior to statistical modeling:
\begin{enumerate}
    \item detecting anomalies: missing data, zero values, blank strings, and infinite numbers (\code{anomalies} function),
    \item drawing and printing univariate distributions of each variable through histograms, bar plots and quantile tables (\code{distributions} function),
    \item drawing plots of variables over time for a specified time variable (\code{timebased} function).
\end{enumerate}
Examples are presented in the readme file in the GitHub repository of the project (\url{https://github.com/sicarul/xray}), but no vignette is attached to it.
Plots\footnote{Access the associated table with \code{archivist::aread("mstaniak/autoEDA-resources/autoEDA-paper/a3a3")}} generated by the package are presented in Figure \ref{fig:xrayplot}.

\subsection{Other packages}\label{other}

As mentioned before, there are numerous R packages that aim to make data exploration faster or the outputs more polished.

For table summaries of data that often include statistical tests, there are a few packages worth mentioning.
The package \CRANpkg{tableone} \citep{tableone} provides a \code{CreateTableOne} function to make publication-ready tables referred to as \textit{Table 1} - traditional name of tables that describe patients' characteristics, usually stratified and including p-values from significance tests.
The \code{describe} function from \CRANpkg{describer} package \citep{describer} prints a summary of a \code{data.frame} or a vector which includes data types, counts and descriptive statistics.
Similarly, the \CRANpkg{skimr} \citep{skimr} package summarises data frames, vectors, and matrices.
It can also handle grouped data frames.
The summary consists of data dimensions, missing and complete value counts, typical descriptive statistics and simple histograms. 
A function of the same name from \CRANpkg{prettyR} \citep{prettyr} returns descriptive statistics for each column in a \code{data.frame}.
This package is focused on improving the aesthetics of R statistical outputs.
Similarly, the package \CRANpkg{Hmisc} \citep{hmisc} includes a \code{describe} function that displays typical descriptive statistics and number of unique and missing values for each column.
The \code{plot} method called on the result of the \code{describe} function returns a dot plot for each categorical and a spike histogram for each continuous column.
The scope of this package is bigger than just Exploratory Data Analysis, as it includes many tools related to regression models.

There are also many packages related to data visualization. 
Two of them are particularly worth mentioning.
The \CRANpkg{ggfortify} package \citep{ggfortify} serves as a uniform interface to plots of different statistical objects, including PCA results that can be used for data exploration and time series plots.
The \CRANpkg{autoplotly} library \citep{autoplotly} was built on top of \pkg{ggfortify} to provide automatically generated, interactive visualizations of many statistical models.
While these two packages are focused on statistical modeling, they can be helpful in exploratory analysis and exemplify the potential of quick and interactive visualization in R.

Two more packages are relevant to our interest. 
\CRANpkg{gpairs} \citep{gpairs} and \CRANpkg{GGally} \citep{ggally} packages implement the generalized pairs plot \citep{gpp}.
This type of plot extends well known scatter plot matrices, that visualize bivariate relationships for many variables, by handling both numerical and categorical variables.
It is helpful in data exploration and shares similarities to \textit{walls of histograms} that can be found in automated EDA libraries.

\section{Feature comparison}\label{chapter:feature}

In this section, we compare how different packages address autoEDA tasks as described in Section \ref{chapter:tasks}.
A quick overview of the functionalities of different packages can be found in Table \ref{comparison}.

\begin{table}[ht]
\centering
\setlength\tabcolsep{3pt}
\begin{tabular}{ll|c|c|c|c|c|c|c|c|c|c|c|c|c|c|c|c|c}
  \toprule
 Task type & Task & a & aE & DE & dM & d & EPD & e & eR & fM & i & R & SE & s & v & x \\ 
  \midrule
 \multirow{4}{*}{Dataset}  & Variable types &  & x & x & x & x &  & x &  & x & x &  & x & x & x &  \\ 
  & Dimensions &  & x & x & x & x & x &  &  & x & x &  & x &  & x &  \\ 
  & Other info &  &  & x &  &  &  &  &  &  & x &  &  &  & x &  \\ 
  & Compare datasets & x &  &  &  &  &  &  &  & x & x &  &  &  & x &  \\ 
  \midrule
 \multirow{5}{*}{Validity} & Missing values &  & x & x & x & x & x & x &  & x & x &  & x & x & x & x \\ 
  & Redundant col. &  & x &  & x & x &  & x &  & x &  &  & x & x & x &  \\ 
  & Outliers &  & x &  & x & x & x &  &  & x &  &  &  &  &  &  \\ 
  & Atypical values &  &  &  & x &  &  &  &  & x &  &  &  &  &  & x \\ 
  & Level encoding &  &  &  & x &  &  &  &  &  &  &  &  &  &  &  \\ 
 \midrule
 \multirow{5}{*}{Univar.} & Descriptive stat. &  & x &  & x & x & x & x &  & x & x & x & x & x &  & x \\ 
  & Histograms &  & x & x & x & x & x & x &  & x & x & x & x & x &  &  \\ 
  & Other dist. plots &  &  & x &  &  & x &  &  &  &  &  &  &  &  &  \\ 
  & Bar plots &  & x & x & x & x & x & x &  & x & x & x & x & x &  &  \\ 
  & QQ plots &  &  & x &  & x &  &  &  &  &  &  & x &  &  &  \\ 
 \midrule
 \multirow{9}{*}{Bivar.} & Descriptive stat. & x &  &  &  & x &  & x &  &  &  & x & x & x &  &  \\ 
  & Correlation matrix &  &  & x &  & x & x &  &  &  & x &  &  &  & x &  \\ 
  & 1 vs each corr. &  & x &  &  &  &  &  & x & x &  &  & x &  &  &  \\ 
  & Time-dependency & x &  &  &  &  & x &  &  &  &  &  &  &  &  & x \\ 
  & Bar plots by target &  & x & x &  & x & x & x &  & x &  & x & x &  &  &  \\ 
  & Num. plots by target &  & x &  &  & x & x & x &  & x &  &  & x &  &  &  \\ 
  & Scatter plots &  &  & x &  &  & x &  & x &  &  & x & x &  &  &  \\ 
  & Contigency tables & x &  &  &  & x &  &  &  &  &  &  & x & x &  &  \\ 
  & Other stats. (factor) &  &  &  &  &  &  &  &  & x & x &  & x &  &  &  \\ 
 \midrule
 \multirow{3}{*}{Multivar.} & PCA &  &  & x &  &  &  &  &  &  &  &  &  &  &  &  \\ 
  & Stat. models & x &  &  &  &  & x & x &  &  &  &  &  &  &  &  \\ 
  & PCP &  &  &  &  &  &  &  &  &  &  &  & x &  &  &  \\ 
 \midrule
 \multirow{6}{*}{Transform.} & Imputation &  &  & x &  & x &  & x &  &  &  &  &  &  &  &  \\ 
  & Scaling &  &  &  &  & x &  &  & x & x &  &  &  &  &  &  \\ 
  & Skewness &  &  &  &  & x &  &  &  &  &  &  &  &  &  &  \\ 
  & Outlier treatment &  &  &  &  & x & x & x &  & x &  &  &  &  &  &  \\ 
  & Binning &  &  & x &  & x &  &  &  & x &  &  &  &  &  &  \\ 
  & Merging levels &  &  & x &  &  &  &  &  & x &  &  &  &  &  &  \\ 
  \midrule
 \multirow{2}{*}{Reporting}  & Reports &  & x & x & x & x &  & x &  &  &  & x & x &  &  &  \\ 
  & Saving outputs & x &  &  &  &  &  &  & x & x &  &  &  & x &  & x \\ 
   \bottomrule
\end{tabular}
\caption{Overview of functionalities of all described packages. Package names were shortened to make the table as compact as possible.
\textbf{a} denotes \pkg{arsenal}, \textbf{aE} - \pkg{autoEDA}, \textbf{DE} - \pkg{DataExplorer}, \textbf{dM} - \pkg{dataMaid}, \textbf{d} - \pkg{dlookr}, \textbf{EPD} - \pkg{ExPanDaR}, \textbf{e} - \pkg{explore}, \textbf{eR} - \pkg{exploreR}, \textbf{fM} - \pkg{funModeling}, \textbf{i} - \pkg{inspectdf}, \textbf{R} - \pkg{RtutoR}, \textbf{SE} - \pkg{SmartEDA}, \textbf{s} - \pkg{summarytools}, \textbf{v} - \pkg{visdat}, \textbf{x} denotes \pkg{xray}.
\textit{Num. plots by target} refers to either histogram, density, violin or box plot.}
\label{comparison}
\end{table}

\subsection{Data description}

Almost all packages contain functions for summarizing datasets. 
Tools that support data validity analysis are less common.

\subsubsection{Whole dataset summaries}

\begin{figure}
    \centering
    \includegraphics[scale=0.6]{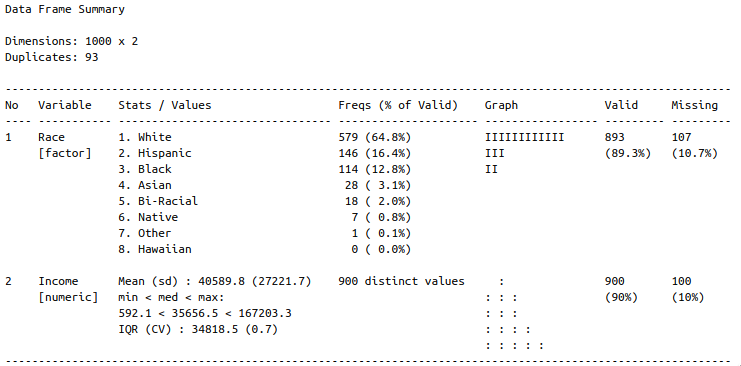}
    \caption{An example of whole data frame description that includes univariate summary and simple graphics. Created with the \code{dfSummary function} (\pkg{summarytools} package v. 0.9.3).}
    \label{fig:wholedfsum}
\end{figure}

Most packages that provide a whole dataset summary take a similar approach and present names and types of variables, number of missing values and sometimes unique values or other statistics. This is true for \pkg{summarytools} (\code{dfSummary} function), \pkg{autoEDA} (\code{dataOverview} function), \pkg{dataMaid} (\code{makeDataReport} result), \pkg{funModeling} (\code{df\_status} function), \pkg{explore} (\code{describe} function), \pkg{ExPanDaR} (\code{prepare\_descriptive\_table} function), and \pkg{DataExplorer} (\code{introduce} function). These outputs are sometimes mixed with univariate summaries. 
That is the case for one of the most popular summary-type functions: the \code{dfSummary} functions from the \pkg{summarytools} package.
An example is given in Figure \ref{fig:wholedfsum}.

In the \pkg{dlookr} package, summaries for numerical variables and categorical variables are only presented separately in the report (\code{describe} function).

The \pkg{visdat} package introduces the most original summaries of the full dataset. 
The drawback of this approach is that it is not well suited for high dimensional data.
But for a smaller number of variables, it gives a good overview of the dataset.

\subsubsection{Data validity}

Some packages can perform automated checks for the data, including at least outlier detection.
The \pkg{dataMaid} package's main purpose is to find inconsistencies and errors in the data.
It finds possible outliers, missing values, low-frequency and possibly miscoded factor levels.
All this information can be summarised in a quality report.
The \pkg{dlookr} package covers similar functionality.
There are two main differences: the report does not describe possibly miscoded factors, but outlier analysis is supplemented with plots showing variable distribution before and after removing the outliers.
In all cases, the analysis is rather simple, for example in zero-inflated variables non-zero values are treated as outliers (\pkg{dlookr}).
The \pkg{ExPanDaR} package handles outliers by providing a function that calculates winsorized or trimmed mean.
Other packages only provide information about the number of missing values/outliers and identify columns that consist of a single value.

\subsection{Data exploration}

While multivariate analysis is rarely supported, there are many tools for descriptive and graphical exploration of uni- and bivariate patterns in the data.

\subsubsection{Univariate statistics}

All the tools that support univariate analysis take a similar approach to univariate analysis.
For categorical variables, counts are reported and bar plots are presented, while histogram or boxplots and typical descriptive statistics (including quantiles, sometimes skewness) are used for continuous variables.

In \pkg{dataMaid} and \pkg{dlookr} packages, these plots are presented variable-by-variable in the report.
In other packages (\pkg{DataExplorer}, \pkg{funModeling}, \pkg{SmartEDA}, \pkg{inspectdf}) groups of plots of the same type are shown together - as a wall of histograms or bar plots.
Similarly, the \pkg{explore} package present all the plots at once.
The \pkg{ExPanDaR} package allows user to choose variables to display in a \pkg{shiny} applications.
Notably, \pkg{dlookr} reports skewness of variables and in case a skewed variable is found, it shows the distribution after some candidate transformations to reduce the skewness have been applied.
This library also reports normality.
The \pkg{SmartEDA} package also reports skewness and displays QQ plots against normal distribution, but it does not provide any means of reducing skewness.

\subsubsection{Bivariate statistics}

\begin{figure}
    \centering
    \includegraphics[scale=0.9]{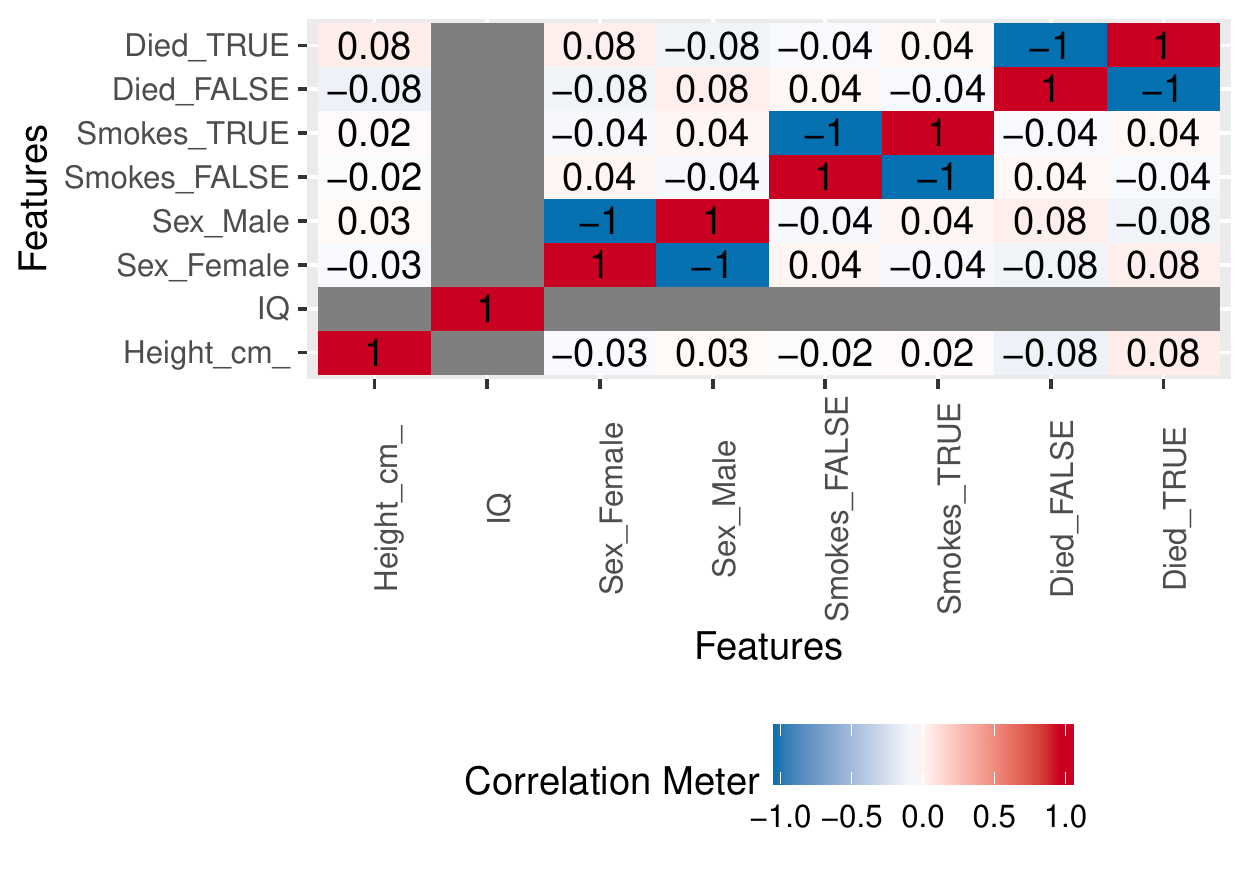}
    \caption{Correlation plot as returned by the \code{DataExplorer::plot\_correlation} function.}
    \label{fig:decorr}
\end{figure}

The \pkg{funModeling} and \pkg{SmartEDA} packages only support calculating correlations between variables and a specified target.
\pkg{DataExplorer} and \pkg{visdat} packages can plot correlation matrices.
They differ in categorical variables treatment. 
Some packages require only numerical features (\pkg{visdat}).
Interestingly, in \pkg{DataExplorer}\footnote{Access the plot with \code{archivist::aread("mstaniak/autoEDA-resources/autoEDA-paper/0526")}}, low-cardinality categorical features are converted to 0-1  variables and plotted alongside numerical variables, as seen in Figure \ref{fig:decorr}.

The \pkg{arsenal} package only presents variable summaries by levels of a chosen categorical variable.
The report from the \pkg{autoEDA} package consists of a limited number of bar plots/boxplots with target variable as one of the dimensions.
Similarly, in \pkg{DataExplorer}, \pkg{dlookr}, \pkg{funModeling} and \pkg{SmartEDA}, scatter plots and box plots or histograms with a specified target variable on one of the axis can be plotted.
Additionally, \pkg{funModeling} and \pkg{dlookr} draw histograms/densities of continuous features by the target.
In \pkg{shiny} applications provided by \pkg{ExPanDaR} and \pkg{explore} packages, the user can choose target variables and explanatory variables to display bivariate plots.
Interestingly, scatter plots provided by the \pkg{ExPanDaR} package can be extended to display multivariate dependencies by mapping variables to the size and color of the points.
The \pkg{funModeling} package also has unique options: drawing bar plots of discretized variables by the target and quantitative analysis for binary outcome based on representativeness and accuracy.
\pkg{arsenal}, \pkg{summarytools} and \pkg{SmartEDA} also feature contingency tables.
Moreover, \pkg{exploreR} and \pkg{ExPanDaR} packages use linear regression plots and statistics to find relationships between the target and other variables.
The \pkg{explore} package can only handle binary targets, but it allows the user to fit and plot a decision tree model.

\subsection{Data cleaning and data transformation} 

The \pkg{dataMaid} package assumes that every decision regarding the data should be made by the analyst and does not provide any tools for data manipulation after diagnosis.
Most of the packages only provide exploration tools.
Exceptions are \pkg{dlookr}, \pkg{funModeling}, \pkg{DataExplorer} and \pkg{exploreR}.
\pkg{DataExplorer} provides tools for normalization, imputation by a constant, merging levels of factors, creating dummy variables and transforming columns.

The \pkg{dlookr} package can create a report that presents different possible transformations of features.
Missing values can be imputed by mean/median/mode and distributions of variables before and after the procedure can be compared.
The same is done for imputation of outliers.
Logarithmic and root square transforms are proposed for skewed variables.
Different methods of binning continuous variables are also presented, including Weight of the Evidence.

The \pkg{funModeling} package can perform discretization of a variable using an equal frequency criterion or gain ratio maximization.
It can also scale variables to the interval $[0, 1]$.
Outliers can be treated using the Tukey or Hampel method.

\subsection{Reporting}

\pkg{DataExplorer}, \pkg{dlookr}, \pkg{dataMaid}, \pkg{SmartEDA}, \pkg{explore} and \pkg{RtutoR} have an option of generating a report and saving it to a file.
Such a report usually consists of all or most possible outputs of the package.
The plots and summaries are organized by the exploration task (for example univariate, then bivariate analysis) and either simply variable-by-variable (\pkg{dataMaid}, \pkg{dlookr}) or grouped by variable type (\pkg{DataExplorer}, \pkg{SmartEDA}).
The \pkg{autoEDA} package generates a minimal report with bivariate plots.
Packages \pkg{arsenal}, \pkg{funModeling}, \pkg{xray}, \pkg{summarytools} and \pkg{exploreR} have an option of saving outputs - plots or tables - to files.

\section{Discussion}\label{chapter:summary}

Automated EDA can be either directed towards a general understanding of a particular dataset or be more model-oriented, serving as a foundation for good modeling.
While presented packages include some tools related to simple variable transformations, they are more focused on data understanding.
For this task, they have many advantages.
In this section, we summarize the strong points of existing tools and point out some possible improvements and new directions for autoEDA.

\subsection{Strengths of autoEDA packages} 

\begin{enumerate}
    \item The packages \pkg{dlookr}, \pkg{dataMaid}, \pkg{DataExplorer}, \pkg{SmartEDA} are capable of creating good quality reports.
    \item \pkg{DataExplorer} has very good visualizations for PCA.
    \item \pkg{DataExplorer} handles categorical variables on correlation plots by creating dummy features, which is a unique idea compared to other packages.
    \item The \pkg{visdat} package, while probably not the best choice for high dimensional data, features interesting take on initial whole dataset exploration.
    \item The \pkg{dlookr} package is capable of selecting skewed variables and proposing transformations. Some of the other packages display binned continuous variables, which can also help in seeing visualizing dependencies.
    \item \pkg{dataMaid} is a good tool for finding problems in the data. Thanks to the structure of \code{check} and \code{summarize} functions results, discovered issues can be treated effectively. 
    \item For datasets with a moderate number of features, \pkg{DataExplorer}, \pkg{funModeling}, \pkg{dlookr} and \pkg{SmartEDA} give a reasonable insight into variables distributions and simple relationships.
    \item \pkg{SmartEDA} package provides a method of visualizing multivariate relationships - parallel coordinate plot.
    \item The \pkg{exploreR} package provides usefuls tool for assessing bivariate relationship through linear regression.
\end{enumerate}
We can see that tasks related to data quality and whole dataset summary are well by the existing libraries. 
Getting the big picture of the data and finding possible data quality problems is easy, especially with the \pkg{dataMaid} package.
For classical applications, for example, statistical analyses in medicine, the current tools provide very good tables, such as the ones from \pkg{tableone} or \pkg{arsenal} packages, and uni-/bivariate plots.
The \pkg{inspectdf} and \pkg{summarytools} packages can also provide quick insights into a dataset.
Univariate analysis can be performed either variable-after-variable (\pkg{dlookr}, \pkg{dataMaid}), where we can see the statistical properties of each variable, or as groups of plots based on variable type (\pkg{DataExplorer}, \pkg{funModeling}).
Both ways can be useful for a reasonable number of predictors.
While multivariate tools are scarce, the available tools, PCA in \pkg{DataExplorer} and PCP in \pkg{SmartEDA}, are very well done.
Notably, the \pkg{ExPanDaR} package provides very high flexibility thanks to the possibility of interactively choosing variables to display, adding new variables on-the-fly and customizing plots in the \pkg{shiny} application.

\subsection{Future directions and possible improvements}

The field of autoEDA is growing.
New packages are being developed rapidly - there are recent additions from April and May. 
Features are added to existing packages and bugs are corrected, as new issues are suggested by users on GitHub. 
At this moment, we can identify the following problems and challenges.

All the presented tools can fail in situations with imperfect data. 
In particular, they are usually not robust to issues like zero-variance/constant variables.
Such problems are expected to be solved in the nearest future, as suggested for example by issues in the GitHub repo of the \pkg{DataExplorer} package.
In general, error messages can be uninformative. 
Moreover, in some situations, they lack flexibility. For example, in \pkg{DataExplorer} arguments can be passed to \code{cor} function, but not to \code{corrplot} function.

In the case of \textit{walls of histograms} (or bar plots), no selection is being done and no specific order is chosen to promote most interesting distributions. The same is true for automatically created reports.  
This problem is only addressed by the \pkg{RtutoR} package, which allows to select top \texttt{k} relevant variables.
Moreover, for high-dimensional data or high-cardinality factors, the plots often become unreadable or impractical. Partial solutions to these problems are applied. For example, \pkg{DataExplorer} removes too large factors from the panels.
More generally, many GitHub issues for the described packages are related to customizing and improving plots and output tables.
It is a challenging task due to the diversity of possible input data and a major concern for developers of autoEDA packages.
    
Typical EDA tasks are limited to exploring bivariate relationships. 
Searching for higher dimensional dependencies would be interesting, for example by adding color and size dimensions to the plots, which was already done in the \pkg{ExPanDaR} package.
For \textit{wall of plots} type of display, such an addition would result in a large number of new plots.
Thus, it would require a proper method of finding the most relevant visualization.
Interactivity partially helps address this issue.    
PCA, parallel coordinate plots and model summaries are supported, but each by a separate package. 
It is evident that there is a shortage of multivariate tools.
Univariate regression models can be plotted by the \pkg{exploreR} package.
The \pkg{explore} package plots decision trees for binary target variables.
In other cases, exploration based on simple statistical models (such as scatter plot smoothing) is not an option.
Using regression models and feature transformations to identify and measure relevant relationships could improve bivariate or multivariate analyses supported by automated EDA.

Regarding variable transformation, only one of the packages addresses the issue of skewed variables. 
Proposing transformations of continuous features other than binning would be helpful and could improve visualizations, for example, scatter plots with skewed variables.
Missing data imputation more advanced than imputing a constant is delegated to other packages, although, it is known that imputation by a constant is usually not the best method of missing values treatment.
Some of the above issues limit the packages' usefulness in iterative work. Though, the comparisons of transform and original features and the possibility of applying discovered transformations to data in \pkg{dlookr} package are steps in the right direction.

Support for time-varying variables and non-classical (not IID) problems such as survival analysis is limited or non-existent. For survival analysis, the automation level is low, but there are two notable tools for summarizing dependencies. First is the recognized package \CRANpkg{survminer} \citep{survminer}, which helps visualize survival curves, while also displaying survival tables and other information. The other tool is the \CRANpkg{cr17} package \citep{cr17}, which includes \code{summarizeCR} function that returns several tables and plots for competing risks analysis. More tools for fast visualization of at least bivariate relationships in such problems would be a big help for analysts.
Cluster analysis is sometimes regarded as a part of the EDA process, but it is not available in any of the packages.

The tools available in R have a similar range to other languages' libraries, for example from Python. 
Python packages such as Dora \citep{dora} or lens \citep{lens} also cover feature-by-feature descriptive statistics and plots, bivariate visualizations of the relationships between predictors and target variable, contingency tables, basic data transformations, and imputation. Tools for visual data exploration supports also tools for visual model exploration like \CRANpkg{DALEX} \citep{DALEX} or \CRANpkg{iml} \citep{iml}. In both cases, visual summaries help to quickly grasp key relations between variables or between input features and model predictions.

Since EDA is both closely connected to feature engineering and based on visual insights, automated EDA can draw from existing tools for automated feature extraction like SAFE ML \citep{2019arXiv190211035G} or TPOT \citep{tpot} and visualization recommendations.
When it comes to aiding visual exploration of a dataset, standalone software carries possibilities beyond what we can expect from R packages or analogous libraries in other languages.
A recent notable example is DIVE \citep{2018-dive}.
It is an example of a growing number of tools for visual data exploration that aim to distinguish between relevant and irrelevant visualization and help the analyst find the most interesting plots.
DIVE is one of the \textit{mixed-initiative visualization systems}, meaning it uses both statistical properties of the dataset and user interactions to find the relevant plots.
Building recommendation systems into autoEDA tools can help address the issue of dealing with high-dimensional data and multivariate dependencies by letting the ML-based system deal with the complexity of a large number of candidate visualizations. 
AI-assisted data exploration can be even faster and more efficient.

As autoEDA tools are still maturing, the efforts in the field are somewhat fragmented.
Many packages try to achieve similar goals, but they can be quite inconsistent.
It is especially visible in the multiplicity of names for the \code{summary}-type function to describe a whole data frame.
As the libraries develop, new standards and conventions should be proposed.

\section{Acknowledgement}

This work was financially supported by the NCN Opus grant 2016/21/B/ST6/02176.

\nocite{*}
\bibliography{staniak-biecek}

\address{Mateusz Staniak \\
  Faculty of Mathematics and Information Science\\
    Warsaw University of Technology\\
    Poland \\
  \email{mtst@mstaniak.pl}}

\address{Przemysław Biecek \\
    Faculty of Mathematics,  
    Informatics and Mechanics \\
    University of Warsaw\\
  Poland \\
  Samsung R\&{D} Institute Poland (SRPOL) \\
ORCiD: 0000-0001-8423-1823\\
  \email{przemyslaw.biecek@gmail.com}}

\end{article}

\end{document}